\documentclass[draft]{agujournal2019}
\usepackage{natbib}
\usepackage{url} %this package should fix any errors with URLs in refs.
\usepackage{lineno}
\usepackage{amsmath,amssymb}
\usepackage{wasysym}
\linenumbers
\newcommand{\lya}{Ly$\alpha$}
\newcommand{\grl}{Geophys. Res. Lett.}
\newcommand{\jgr}{J. Geophys. Res.}
\newcommand{\solphys}{Solar Phys.}
\newcommand{\apj}{Astrophys. J.}
\newcommand{\apjl}{Astrophys. J. Letters}
\newcommand{\apjs}{Astrophys. Journal Supp.}
\newcommand{\aap}{Astronomy \& Astrophysics}
\newcommand{\procspie}{Proc. SPIE}
\draftfalse

\journalname{Space Weather}

\begin{document}
\nolinenumbers

%% ------------------------------------------------------------------------ %%
%  Title
%
% (A title should be specific, informative, and brief. Use
% abbreviations only if they are defined in the abstract. Titles that
% start with general keywords then specific terms are optimized in
% searches)
%
%% ------------------------------------------------------------------------ %%

% Example: \title{This is a test title}

%\title{Statistical Study of Major Solar Flares Observed in Lyman-alpha Emission During Solar Cycle 24 Using GOES-15/EUVS-E}
\title{Lyman-alpha Variability During Solar Flares Over Solar Cycle 24 Using GOES-15/EUVS-E}

%% ------------------------------------------------------------------------ %%
%
%  AUTHORS AND AFFILIATIONS
%
%% ------------------------------------------------------------------------ %%

% Authors are individuals who have significantly contributed to the
% research and preparation of the article. Group authors are allowed, if
% each author in the group is separately identified in an appendix.)

% List authors by first name or initial followed by last name and
% separated by commas. Use \affil{} to number affiliations, and
% \thanks{} for author notes.
% Additional author notes should be indicated with \thanks{} (for
% example, for current addresses).

% Example: \authors{A. B. Author\affil{1}\thanks{Current address, Antartica}, B. C. Author\affil{2,3}, and D. E.
% Author\affil{3,4}\thanks{Also funded by Monsanto.}}

\authors{Ryan O. Milligan\affil{1,2,3}, Hugh S. Hudson\affil{2,4}, Phillip C. Chamberlin\affil{5}, Iain G. Hannah\affil{2}, Laura A. Hayes\affil{3}}

% \affiliation{1}{First Affiliation}
% \affiliation{2}{Second Affiliation}
% \affiliation{3}{Third Affiliation}
% \affiliation{4}{Fourth Affiliation}

\affiliation{1}{Astrophysics Research Center, School of Maths and Physics, Queen's University Belfast, UK BT7 1NN}
\affiliation{2}{SUPA School of Physics \& Astronomy, University of Glasgow, UK G12 8QQ}
\affiliation{3}{NASA Goddard Space Flight Center, Greenbelt Road, Greenbelt, MD USA, 20771}
\affiliation{4}{Space Science Lab., University of California, Berkeley}
\affiliation{5}{Laboratory for Atmospheric and Space Physics, Boulder, CO.}
%(repeat as many times as is necessary)

%% Corresponding Author:
% Corresponding author mailing address and e-mail address:

% (include name and email addresses of the corresponding author.  More
% than one corresponding author is allowed in this LaTeX file and for
% publication; but only one corresponding author is allowed in our
% editorial system.)

% Example: \correspondingauthor{First and Last Name}{email@address.edu}

\correspondingauthor{Ryan Milligan}{r.milligan@qub.ac.uk}

%% Keypoints, final entry on title page.

%  List up to three key points (at least one is required)
%  Key Points summarize the main points and conclusions of the article
%  Each must be 100 characters or less with no special characters or punctuation

% Example:
% \begin{keypoints}
% \item	List up to three key points (at least one is required)
% \item	Key Points summarize the main points and conclusions of the article
% \item	Each must be 100 characters or less with no special characters or punctuation
% \end{keypoints}

\begin{keypoints}
\item Lyman-alpha irradiance variability during solar flares can be comparable to that due to solar rotation, only on much shorter timescales.
\item The Lyman-alpha irradiance of a flare exhibits significant anisotropy (center-to-limb-variation).
\item Enhanced Lyman-alpha in the flare impulsive phase appears to be capable of inducing currents in the ionospheric E-layer.
\end{keypoints}

%% ------------------------------------------------------------------------ %%
%
%  ABSTRACT
%
% A good abstract will begin with a short description of the problem
% being addressed, briefly describe the new data or analyses, then
% briefly states the main conclusion(s) and how they are supported and
% uncertainties.
%% ------------------------------------------------------------------------ %%

%% \begin{abstract} starts the second page

\begin{abstract}
The chromospheric Lyman-alpha line of neutral hydrogen (\lya; 1216\AA) is the strongest emission line in the solar spectrum. Fluctuations in \lya\ are known to drive changes in planetary atmospheres, although few instruments have had the ability to capture rapid \lya\ enhancements during solar flares. In this paper we describe flare-associated emissions via a statistical study of 477 M- and X-class flares as observed by the EUV Sensor on board the 15th Geostationary Operational Environmental Satellite, which has been monitoring the full-disk solar \lya\ irradiance on 10~s timescales over the course of Solar Cycle 24. The vast majority (95\%) of these flares produced \lya\ enhancements of 10\% or less above background levels, with a maximum increase of $\sim$30\%. The irradiance in \lya\ was found to exceed that of the 1-8 \AA\ X-ray irradiance by as much as two orders of magnitude in some cases, although flares that occurred closer to the solar limb were found to exhibit less of a \lya\ enhancement. This center-to-limb variation was verified through a joint, stereoscopic observation of an X-class flare that appeared near the limb as viewed from Earth, but close to disk center as viewed by the MAVEN spacecraft in orbit around Mars. The frequency distribution of peak \lya\ was found to have a power-law slope of $2.8\pm0.27$. We also show that increased \lya\ flux is closely correlated with induced currents in the ionospheric E-layer through the detection of the solar flare effect as observed by the Kakioka magnetometer.
\end{abstract}

%% ------------------------------------------------------------------------ %%
%
%  TEXT
%
%% ------------------------------------------------------------------------ %%

%%% Suggested section heads:
% \section{Introduction}
%
% The main text should start with an introduction. Except for short
% manuscripts (such as comments and replies), the text should be divided
% into sections, each with its own heading.

% Headings should be sentence fragments and do not begin with a
% lowercase letter or number. Examples of good headings are:

% \section{Materials and Methods}
% Here is text on Materials and Methods.
%
% \subsection{A descriptive heading about methods}
% More about Methods.
%
% \section{Data} (Or section title might be a descriptive heading about data)
%
% \section{Results} (Or section title might be a descriptive heading about the
% results)
%
% \section{Conclusions}

\section{Introduction}
\label{sec:intro}

The Lyman-alpha (Ly$\alpha$; 1216\AA) line of neutral hydrogen, resulting from the 2p--1s transition, is the brightest emission line in the solar spectrum. During quiescent solar conditions, the wings of the line are formed in mid-chromosphere ($\sim$6,000~K), while the core is formed higher up at the base of the transition region ($\sim$40,000~K). Due to the high abundance of neutral hydrogen in the solar chromosphere, the \lya\ line is optically thick with a broad central reversal. During solar flares nonthermal electrons deposit large excess energies in the chromospheric plasma, generating localized heating and ionization at the flare footpoints. This results in enhanced \lya\ emission although to date there have been very few \lya\ flare profiles recorded at high spectral resolution. Further \lya\ flare emission must come from the corona, where according to the well-established scenario the hot, dense flare plasma cools down from X-ray temperatures and eventually drains back to the lower atmosphere as coronal rain. However, given that the electron density and the abundance of neutral hydrogen are considerably higher in the chromosphere than the corona, it is assumed that the bulk of the \lya\ emission discussed in this paper originates at the flare footpoints \citep{cham18}.

\lya\ photons drive the molecular dissociation of oxygen (O$_2$) in the Earth's mesosphere, allowing ozone (O$_3$) to form, while the photoionization of nitric oxide (NO) leads to the formation of the dayside ionospheric D-layer ($\sim$60--90~km; \citealt{chub57,lean85,wood95}). Changes in the Sun's output at these wavelengths can therefore have significant implications for the dynamics and composition of the terrestrial environment. Several studies have reported on variations in \lya\ due to solar rotation and over the course of solar cycles \citep{lean83,lile08,wood08}. \cite{wood00} reported that the mean variability of \lya\ due to the 27-day solar rotation across Solar Cycles 18--22 was 9\%, dropping to 5\% at solar minimum and increasing to 11\% at solar maximum. Solar \lya\ variability over the course of a solar cycle varied between a factor of 1.5 and 2.1. 

The importance of \lya\ emission from other stars is also being realized as the search for habitable exoplanets intensifies. However, stellar \lya\ is almost impossible to detect due to extinction by the interstellar medium, and so indirect reconstruction techniques need to be applied \citep[e.g.,][]{lins13}. Furthermore, the integration times required to measure stellar EUV emission can be as much as 10$^3$~s, which would make the detection of any flare-related variations in \lya\ very difficult \citep{chri03}.

%Changes in \lya\ irradiance during solar flares, 
Solar variability in \lya\ due to solar flares has not been extensively studied, largely due to instrumental limitations and their duty cycles. \cite{canf80} were among the first to publish temporal variations in the \lya\ line profile during two solar flares using the NRL spectrograph as part of the Apollo Telescope Mount onboard Skylab. \cite{brek96} reported a 6\% increase in \lya\ irradiance using Upper Atmosphere Research Satellite/Solar-Stellar Irradiance Comparison Experiment (UARS/SOLSTICE), while \cite{wood04} reported a 20\% increase in the line core and a factor of two increase in the line wings during the famous 28 October 2003 X28 flare from a serendipitous Solar Radiation and Climate Experiment (SORCE) SOLSTICE \citep{mccl05} observation. While both SOLSTICE instruments capture disk-integrated emission, their cadence is not sufficient for capturing rapid temporal variations during flares. The Large Yield Radiometer (LYRA; \citealt{hoch06,domi13}) instrument on Project for On-Board Autonomy-2 (PROBA2) also captures full-disk \lya\ emission but with very high cadence (50~ms). \cite{kret13} reported \lya\ signatures in 11 flares (C- and M-class) using LYRA, with only a 0.6\% increase in emission detected during a detailed study of an M2 flare. The authors comment that this unusually low contrast may be due to the severe detector degradation suffered by the \lya\ channel, but did not give a detailed explanation. \cite{raul13} also reported a $<$1\% increase in \lya\ emission during seven solar flares using LYRA, but they also did not detect any appreciable ionospheric disturbances. 

As \lya\ is formed in the chromosphere, its temporal variability during flares should mimic that of the nonthermal electrons responsible for chromospheric heating (i.e. impulsive hard X-ray (HXR) bursts). \cite{kret13} discussed how \lya\ flare emission from LYRA varied more gradually than expected for an impulsively heated chromospheric feature. A similar discrepancy was reported by \cite{mill16} for the Extreme-ultraviolet Variability Experiment (EVE; \citealt{wood12}) Multiple EUV Grazing Spectrograph Photometer (MEGS-P) on the Solar Dynamics Observatory (SDO; \citealt{pesn12}). In the case of EVE, this was eventually attributed to a Kalman filter used to smooth the data during processing; this was ultimately replaced with a Fourier transform filter (Don Woodraska; Private Communication, 2017). \cite{mill16} also showed that the \lya\ lightcurves from the EUV Sensor (EUVS; \citealt{vier07,evan10}) on {\it Geostationary Operational Environmental Satellite (GOES)-15} behaved much more impulsively, as expected, consistent with Lyman continuum (LyC) observations from SDO/EVE MEGS-B and with the temporal behaviour of the \lya\ emission from the flare presented by \cite{wood04}. In a statistical study of \lya\ flares \cite{kret15} derived a scaling relationship between the \lya\ fluence and soft X-ray (SXR) fluence for around 100 M- and X-class flares, using 1-minute averaged GOES-15/EUVS-E data. He found that \lya\ was around an order of magnitude stronger than X-rays, with a large degree of scatter that was attributed to either geocoronal absorption or inherent complexities of the \lya\ line formation.

As well as the space weather implications of changes in \lya\ emission, \citet{mill14} showed that some 6--8\% of the energy deposited in the chromosphere by nonthermal electrons can be radiated away by the \lya\ line alone (10$^{30}$~erg; see also \citealt{mill12}). This single emission line therefore becomes one of the most important observables for studies of flare energetics. \nocite{daco09} Rubio da Costa \textit{et al.} (2009) reached a similar conclusion ($<$10\% of nonthermal energy) using \lya\ images from the Transition Region and Coronal Explorer (TRACE; \citealt{hand99}) \cite[see also][]{nusi06}. \cite{mill17} also showed that the chromosphere responds dynamically to an impulsive disturbance at its acoustic cutoff frequency, as evidenced by 3-minute oscillations in the \lya\ time profile from GOES-15/EUVS-E, which appeared in phase with similar periodicities measured in the LyC from SDO/EVE and the 1600\AA\ and 1700\AA\ lightcurves from the Atmospheric Imaging Assembly (AIA; \citealt{leme11}) also on SDO. Thus some of a flare's energy could be dispersed via wave damping (kinetic energy), leading to different radiation signatures.

This paper presents a self-consistent, statistical analysis of almost 500 major solar flares observed in \lya\ emission. The GOES-15/EUVS-E data have clean time profiles, with high signal-to-noise ratios for major events, making quantitative analysis possible. The time series abundantly confirm the recent discovery of an impulsive-phase signature for \lya, akin to that seen in HXR. This opens a new quantitative channel for modeling of the crucially important flare processes at and near the footpoints of coronal magnetic structures. We primarily present a catalog of \lya\ energetics, frequency distributions, and center-to-limb variations for future solar-terrestrial studies, and to highlight the availability and importance of an under-utilized dataset. Section~\ref{sec:data_anal} describes the data analysis techniques, including accounting for the effects of geocoronal absorption. The findings are presented in Section~\ref{sec:results}, including a section on ionospheric and geomagetic consequences. The conclusions are presented in Section~\ref{sec:conc}.

\section{Data Selection and Reduction}
\label{sec:data_anal}

The GOES series of spacecraft have been providing a near-uninterrupted measurement of the solar X-ray irradiance since 1975, beginning with GOES-1, via their X-Ray Sensor (XRS) instruments \citep{hans96}. These data have become the ``industry standard'' for classifying solar flare magnitudes from the peak of the 1-minute averaged 1--8\AA\ channel. After the launch of GOES-8 in 1994, the reported X-rays fluxes were found to differ from those of the earlier GOES satellites due to more accurate calibration. The National Oceanic and Atmospheric Association (NOAA, who provide the data) decided to scale all XRS data from GOES-8 onwards to match those of earlier missions, in order to preserve the long-term record. To recover the ``true'' X-ray fluxes, the long-channel data (1--8\AA) ought to be divided by 0.7, while the short-channel data (0.5--4\AA) needs to be divided by 0.85. This correction has not been applied in this study in order to allow comparisons with previous works, although it is factored in to the GOES software in SSWIDL (SolarSoftWare Interactive Data Language; \citealt{free98}) when computing the total thermal energy since the temperature response functions are based on the corrected data (Kim Tolbert; Private Communication 2019; \url{https://hesperia.gsfc.nasa.gov/rhessidatacenter/complementary_data/goes.html}). Only XRS data from GOES-15, with a cadence of 2~s, have been used in this study.
%Nevertheless, the corresponding $\sim$40\% increase in 1--8\AA\ flux would have little effect on the correlations with \lya\ emission. 

The launch of GOES-13 in 2006 (and subsequently, GOES-14 and -15) saw the inclusion of the EUVS in order to characterize the solar EUV irradiance with a cadence of 10.24~s. EUVS comprises five channels: A, B, C, D, and E, covering the 50--170\AA, 240--340\AA, 200--620\AA, 200--800\AA, and 1180--1250\AA\ wavelength ranges, respectively \citep{vier07}, with the E-channel centered on the \lya\ line at 1216\AA. The E-channel data have been converted to irradiance by scaling to a \textit{Whole Heliosphere Interval} quiet-Sun reference spectrum (\url{http://lasp.colorado.edu/lisird/data/whi_ref_spectra}; \citealt{wood09}). The conversion to physical units for irradiance will therefore not reflect flare-related time variations of the line profile, generating some systematic uncertainties. Over time, the EUVS-E suffers from degradation, which is compensated for by scaling the daily average values to those from SORCE/SOLSTICE. While \lya\ measurements from GOES-13 and -14 have been sporadic, GOES-15 has provided continuous coverage since its launch in 2010. GOES-15 continues to take measurements in the EUV although at the time of writing, only data up to 6 June 2016 have been made publicly available (\url{https://www.ngdc.noaa.gov/stp/satellite/goes/doc/GOES_NOP_EUV_readme.pdf}). Only Version 4 of the 10.24~s cadence EUVS-E data from GOES-15 were used in this study.

\begin{figure}[!t]
\includegraphics[width=\textwidth]{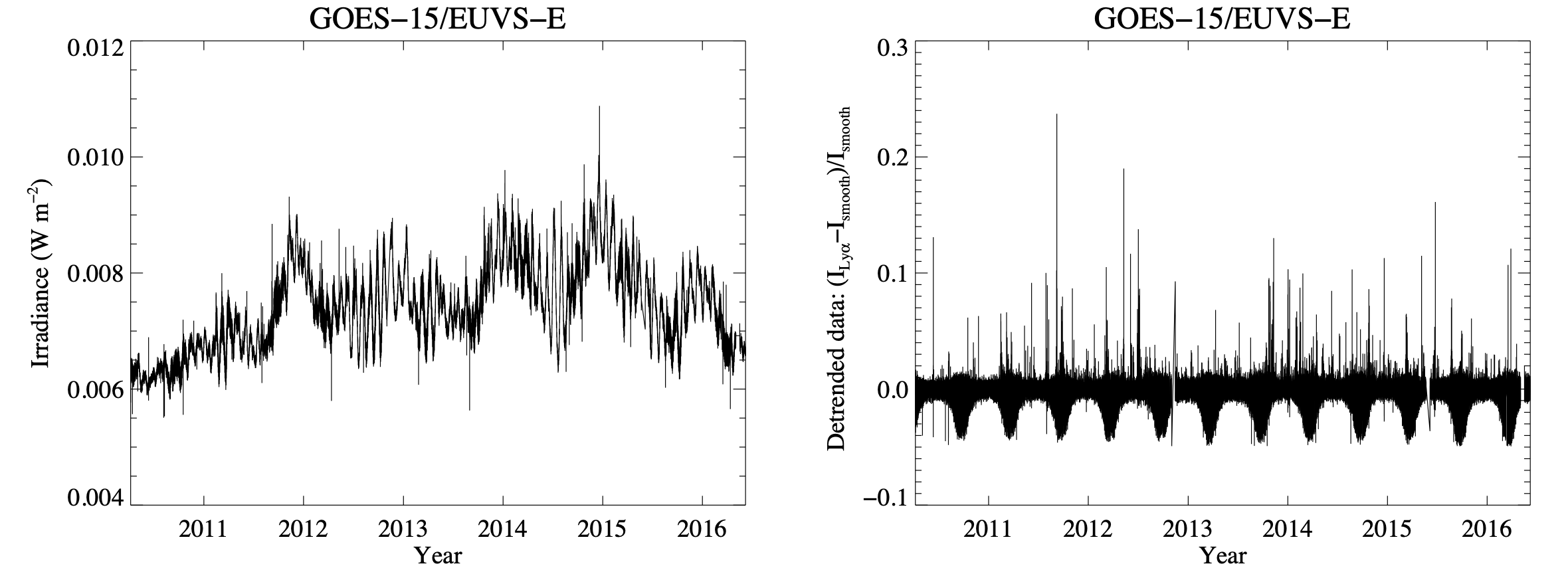}
\caption{Left panel: Plot of all the GOES-15/EUVS-E data currently available, from 7 April 2010 until 6 June 2016, binned by a factor of 12 ($\sim$2~minutes) for display purposes. The 27-day active region period is clearly visible. Right panel: The same six years of \lya\ data, detrended using a 120-point ($\sim$20-minute) smoothing function and normalized. This illustrates the flaring epochs and biannual eclipse seasons.}
\label{fig:lya_solar_cycle}
\end{figure}

 The SDO/EVE instrument has a MEGS-B component, which returns whole-Sun spectra at EUV wavelengths (330-1050\AA; \citealt{crot07}) with 10~s time resolution and 1\AA~spectral resolution. The same instrument measures spatially- and spectrally-integrated \lya\ emission via its MEGS-P photometer, also at 10~s cadence. Unforeseen detector degradation post-launch meant that MEGS-B and -P were only able to observe the Sun in \lya\ for 3 hours per day, plus 5 minutes every hour for most of the SDO mission, and so it has observed significantly fewer flares than GOES-15 (see Table~\ref{tab:db_flares}). In 2015, the flight software was updated to have MEGS-P (and MEGS-B) respond to increases in SXR emission detected by its ESP channel, making EVE a dedicated flare instrument. Recently (19 April 2018), to further preserve the signal-to-noise ratio as the instrument deteriorates, the cadence was reduced to 60~s. 

\begin{table}
\begin{center}
\caption{Breakdown of the number of flares used in this study and comparisons with other datasets over the same six year time period.}   
\begin{tabular}{lccc}
\hline
Database			&M-class	&X-class	&Total	\\
\hline
NOAA/GOES		    &677		&45		&722		\\
SSW Latest Events	&573		&33		&606		\\
This study			&446		&31		&477		\\
SDO/EVE MEGS-P	    &94		    &8		&102	    \\
\hline
\end{tabular}
\label{tab:db_flares}
\end{center}
\end{table} 

\begin{figure}[!b]
\begin{center}
\includegraphics[width=\textwidth]{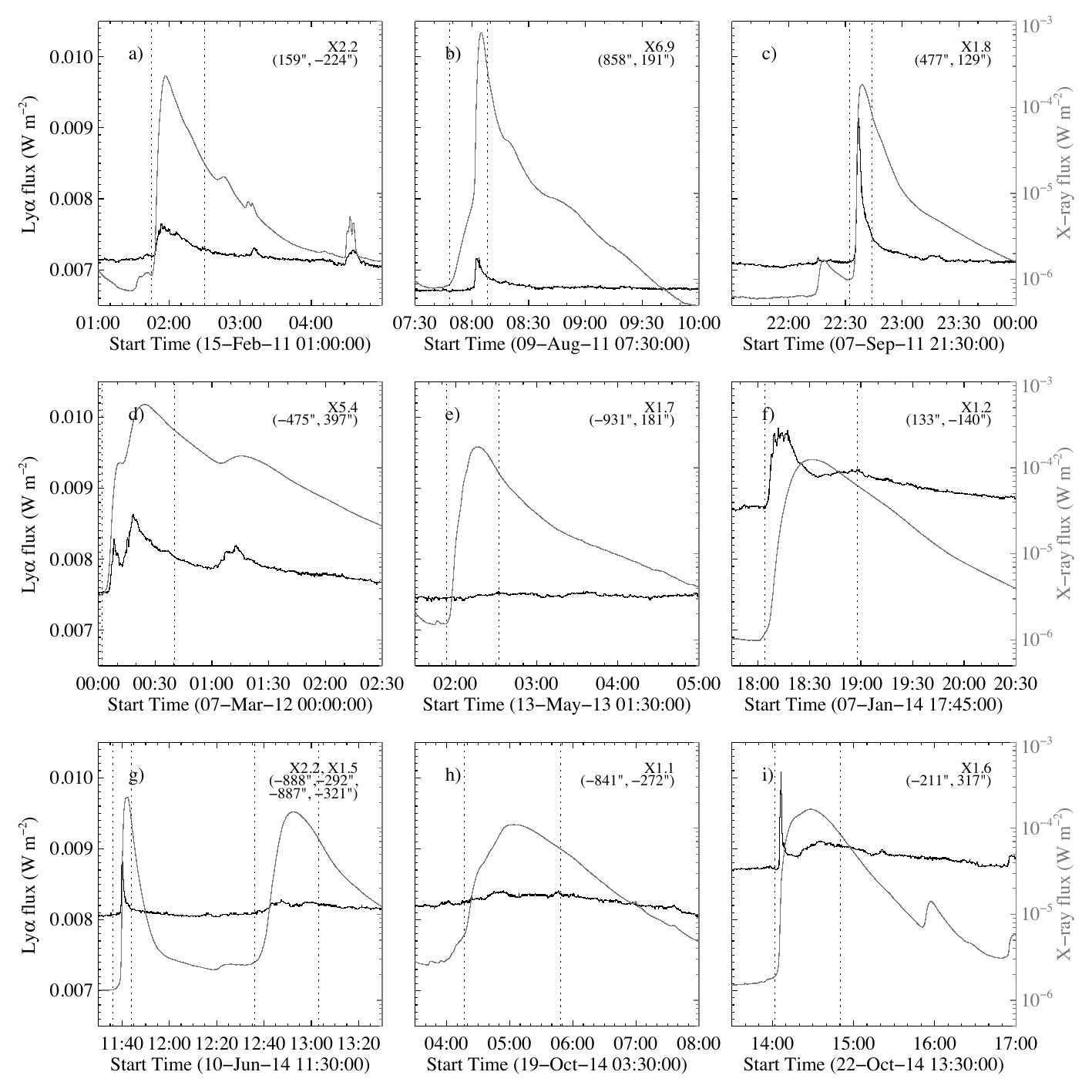}
\caption{Selection of X-class flares shown in both \lya\ (black) and SXR (grey). GOES class and flare locations derived from SDO/AIA are annotated. The vertical dotted lines on each panel denote the (GOES/XRS) start and end time of each X-class flare as listed in the NOAA flare catalog.}
\label{fig:mplot_lya_sxr}
\end{center}
\end{figure}

The left-hand panel of Figure~\ref{fig:lya_solar_cycle} shows the complete six years of currently available GOES-15/EUVS-E data covering Solar Cycle 24. The presence of the 27-day period due to active region rotation is clearly visible. By detrending these data (using a 120-point/20~minute smoothing function) to remove solar-cycle and rotation timescales, the biannual eclipse seasons - seen as dips in the time profile in the right-hand panel of Figure~\ref{fig:lya_solar_cycle} - become apparent. These result from geocoronal absorption by the Earth's uppermost atmosphere, which has substantial opacity to \lya\ line-core photons \citep{meie70,bali19}. Note that the timing and depth of this dip varies over the course of the year as the EUVS instrument peers through different column depths of the geocorona throughout its orbit. The daily dips typically last for around $\pm$4 hours of local midnight, and range from 0.3--6\%, with the greatest decreases occurring at the time of the equinoxes. The geocorona is transparent to X-rays, but there are periods when the GOES satellites are in eclipse, blocking out both X-rays and \lya.

One of the best-established solar flare catalogs available is that from the GOES/XRS photometers, as characterized by the flux of its 1--8\AA\ passband (\url{https://www.ngdc.noaa.gov/stp/solar/solarflares.html}). This list is compiled by NOAA and extends back to 1977. Over the six years of available GOES-15/EUVS-E data, the catalog lists the start, peak, and end times of 677 M-class flares and 45 X-class (see Table~\ref{tab:db_flares}). While the NOAA event list does not always provide heliographic locations for each of these flares, \cite{mill18} recently compiled a list of flares observed by multiple instruments during Solar Cycle 24, which included flare locations based on SDO/AIA 94\AA\ ($\sim$10~MK) difference images. These (coronal) locations were extracted from the SSW Latest Events list, which is accessible through the Heliophysics Events Knowledgebase (HEK; \url{https://www.lmsal.com/hek/}). In this study the locations were needed to derive the center-to-limb variation (CLV) of (excess) \lya\ emission (Section~\ref{sec:clv}). However, this database is incomplete and is missing several months of information. Location information was available for 573 M-class and 33 X-class flare in total. Given that the effect of geocoronal absorption on the flare excess emission was assumed to be nonlinear, and therefore difficult to correct for, all events for which the GOES start and/or end time lay within $\pm$2$\sigma$ of the minima of all geocoronal dips were omitted. This left a sample of 446 M-class flares and 31 X's (477 in total). For comparison, from the launch of SDO until 6 June 2016, EVE MEGS-P observed 94 M-class flares and eight X-classes flare in their entirety (between GOES start and end times), not counting those that were partially observing during its 5-minute observing periods.

Figure~\ref{fig:mplot_lya_sxr} shows a sample of ten X-class flares in both X-rays (grey curves) and \lya\ (black curves) to illustrate the quality of the EUVS-E data and to show the variety of \lya\ responses during flares of comparable X-ray magnitudes. Their GOES classifications and heliographic locations are annotated in the top right corner of each panel. Panel $a$ shows the first X-class flare of Solar Cycle 24 as previously reported by \cite{mill12,mill14}, \cite{mill16}, and \cite{mill17}. Interestingly, the flare in panel $b$ is the largest in the sample (X6.9), but only shows a modest \lya\ increase, while the more common X1.8 flare in panel $c$ exhibited the largest \lya\ contrast (29\% increase) of all the flares studied. Some events displayed multiple \lya\ bursts (panels $d$, $f$, and $i$), while others showed no variation due to being either behind (panel $e$) or close to (panel $h$) the solar limb. Even back-to-back events of similar magnitudes from the same active region can exhibit substantially different \lya\ profiles (panel $g$). 

\subsection{Background Subtraction}
\label{sec:bgsub}

\begin{figure}[!t]
\begin{center}
\includegraphics[width=\textwidth]{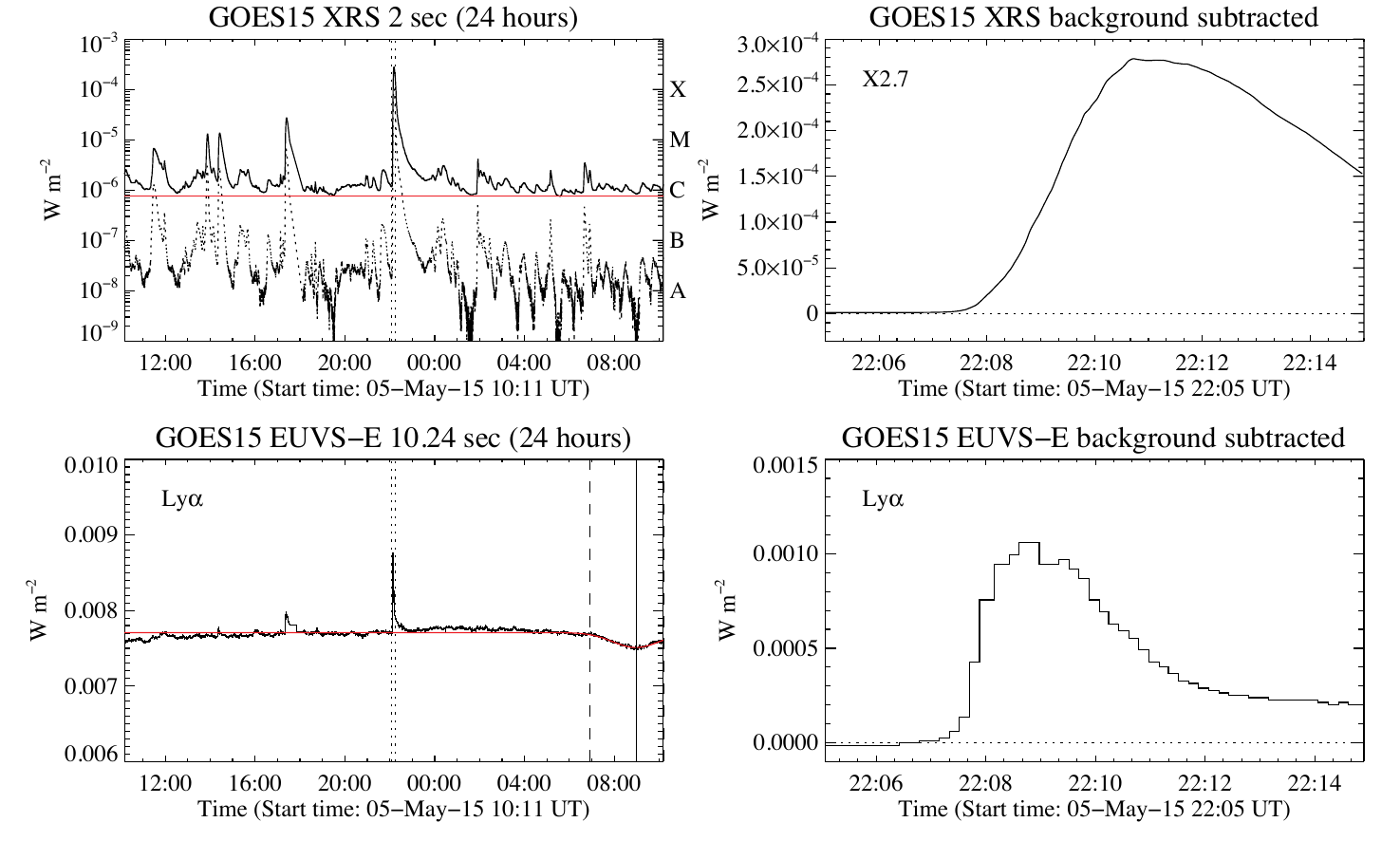}
\caption{Top left panel: 24 hours of GOES-15/XRS data (both channels: 1--8\AA\ and 0.5--4\AA) centered on the peak of the X2.7 flare that occurred on 5 May 2015 (SOL2015-05-05) beginning at 22:05~UT. The horizontal red line denotes the background for this period, taken to be the minimum value in the 1--8\AA\ channel. The two vertical dotted lines denote the beginning and the end of the X2.7 flare as listed in the NOAA event list. Top right panel: The background subtracted GOES/XRS data between the start and end of the GOES event. Bottom left panel: The GOES-15/EUVS-E (\lya) data over the same 24 hour period. The red line marks the background value over this period, taken as a constant equal to the modal value plus a Gaussian to account for geocoronal absorption, seen as a dip in the data around 09:00~UT on 6 May 2015. The vertical solid line marks the minimum of the depth, while the vertical dashed line marks the 2$\sigma$ width. Bottom right panel: background-subtracted \lya\ lightcurve between the start and end of the GOES event.}
\label{fig:xrs_lya_flare}
\end{center}
\end{figure}

Figure~\ref{fig:xrs_lya_flare} illustrates how background subtracted X-ray and \lya\ profiles were generated for all flares considered. The top left panel shows XRS data over a 24-hour period centered on the peak of the X2.7 flare that occurred on 5 May 2015 (SOL2015-05-05). For each flare, the background of the 1--8\AA\ channel was taken to be the minimum value within this 24-hour period (denoted by the horizontal red line). The resulting background subtracted profile, between the start and end times of the GOES event (shown as vertical dotted lines in the two left hand panels), is then shown in the top right hand panel. Note that because the largest flares often have peak fluxes several orders of magnitude above the background level, they are not particularly sensitive to background subtraction \citep{ryan12}. 
%Several other, weaker flares also occurred on this same day. In order to quantify the flare excess emission for each event, the solar background, which itself varies on various timescales, needed to be accounted for. 

The bottom left-hand panel of Figure~\ref{fig:xrs_lya_flare} shows the corresponding EUVS-E \lya\ lightcurve over the same 24-hour period. Comparing with the panel above, it is clear that only the largest flares produce enhanced \lya\ emission above the background level. Also visible is the daily geocoronal absorption dip beginning around 07:00~UT (vertical dashed line). The presence of these geocoronal dips coupled with the fact that changes in \lya\ due to flares are not as substantial as those in X-rays requires a more careful subtraction of the solar background. The entire 24 hour period was fit with an inverted Gaussian to account for the geocoronal dip, plus a constant equal to the modal value over the time interval. Subtracting this background reveals the flare excess emission between the start and end times of the corresponding GOES event (bottom right hand panel). The ratio of the peak \lya\ flux relative to the background value then defines the contrast (percentage increase above background; see Section~\ref{sec:flare_cont}). 

\subsection{Flare Energetics}
\label{sec:energetics}

Having established the flare excess emission in both \lya\ and X-rays between the start and end of each GOES event, the total energy emitted over each wavelength range can be determined by integrating each profile over time. We convert from W~m$^{-2}$ to erg~s$^{-1}$; 1~W~m$^{-2}$ = 2 $\pi$ (1 AU)$^2\times$10$^7$~erg~s$^{-1}$=1.406$\times$10$^{30}$~erg~s$^{-1}$. Deviations from 1 AU as the Earth - and therefore, GOES-15 - completes its orbit around the Sun were also accounted for. The flux ratio from the two XRS channels also allows the derivation of a temperature ($T$) and emission measure ($EM$) of the X-ray emitting plasma via the procedure outlined in \cite{garc94} and \cite{whit05}. Combining these results with the radiative loss function allows the radiative loss rate by the total SXR emitting plasma to be calculated through the simple expression $3n_ek_BTV$, where EM = $n_e^2V$, $n_e$ is the electon density, $V$ is the volume, and $k_B$ is the Boltzmann constant (see Section~\ref{sec:flare_energy}). Coronal abundances and $n_e$=10$^{10}$~cm$^{-3}$ are assumed by default.
%To get the total integrated energy meant also taking the instrumental cadence time into consideration. 

\begin{figure}[!t]
\begin{center}
\includegraphics[width=\textwidth]{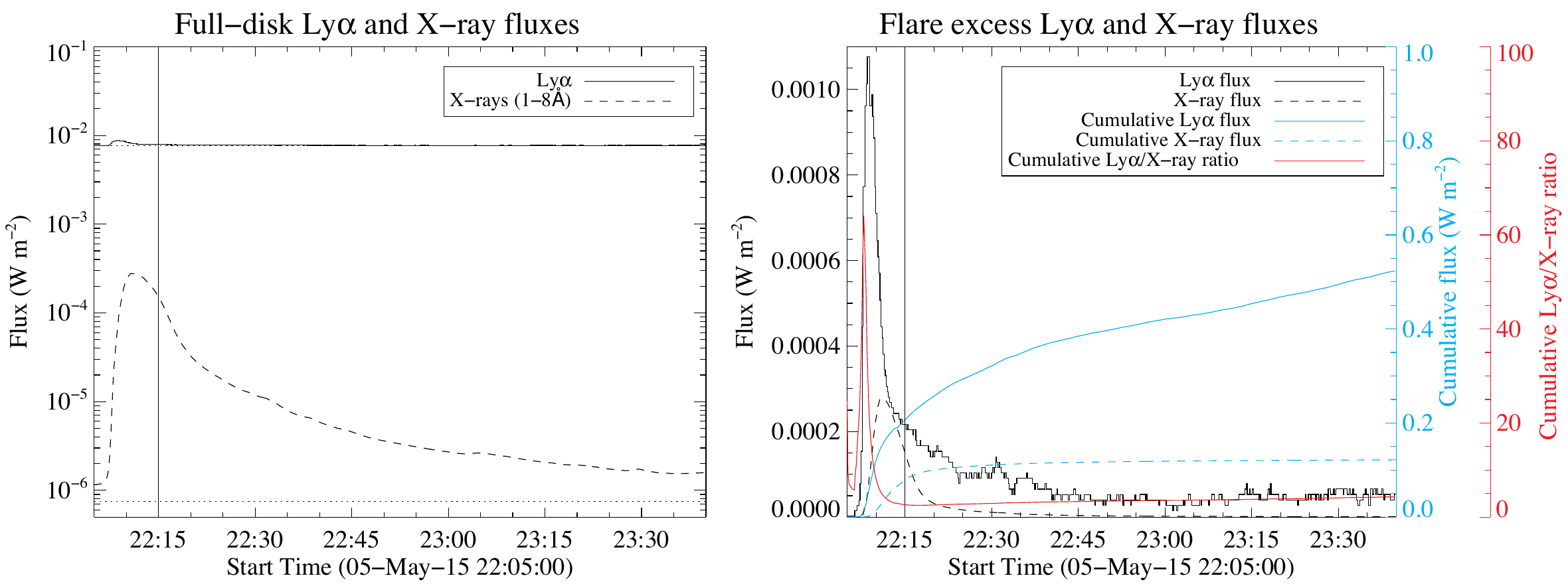}
\caption{Plot of the full-disk \lya\ (solid black curve) and X-ray (dashed black curve) fluxes (left; logarithmic scale) and the flare excess (right; linear scale) during the 5 May 2015 X-class flare. In the left panel the background values are shown as horizontal dotted lines. In the right panel, the cumulative fluxes are plotted as solid and dashed cyan lines, respectively. The red curve denotes the ratio of the cumulative fluxes over time, while the vertical solid black line in each panel marks the end time of the flare as listed in the NOAA flare catalog.}
\label{fig:goes_end_time}
\end{center}
\end{figure}

The biggest uncertainty in the determination of the total energy radiated by each mechanism is that of the end time of the flare as specified in the GOES event list. NOAA stipulate this time to be that when the X-ray flux equals half that of the peak value (\url{https://www.ngdc.noaa.gov/stp/solar/solarflares.html}). In the largest events, X-rays can remain elevated above the background levels for several hours beyond the cataloged end time, while the impulsive nature of \lya\ often sees it approach background levels much more rapidly. Accounting for the `real' flare end time would require fitting the decaying emission with an uncertain approximation, say an exponential function, and extrapolating in time. This is beyond the scope of this paper (see also \citealt{ryan16}). Therefore, to assess the effect of a longer flare duration on the relative energy radiated, the end time of the X-class flare shown in Figure~\ref{fig:xrs_lya_flare} was arbitrarily extended by 90 minutes. As shown in the left hand panel of Figure~\ref{fig:goes_end_time}, the X-ray flux (dashed curve) continued to decay beyond the listed end time (vertical solid line), while the \lya\ flux (solid curve) had almost returned to pre-flare levels. The cumulative radiated energy by the two processes are shown in the right hand panel as cyan curves, along with the background-subtracted lightcurves (black curves). The ratio of the two energies is plotted as a solid red curve, and does not appear to change appreciably between the listed GOES end time and 90 minutes later. The energies for both X-rays and \lya\ reported in Section~\ref{sec:results} can therefore be considered lower limits, although the ratio of the two may be broadly considered independent of the total integration time. 

\section{Results}
\label{sec:results}

\subsection{Flare Contrast}
\label{sec:flare_cont}

\begin{figure}[!t]
\includegraphics[width=\textwidth]{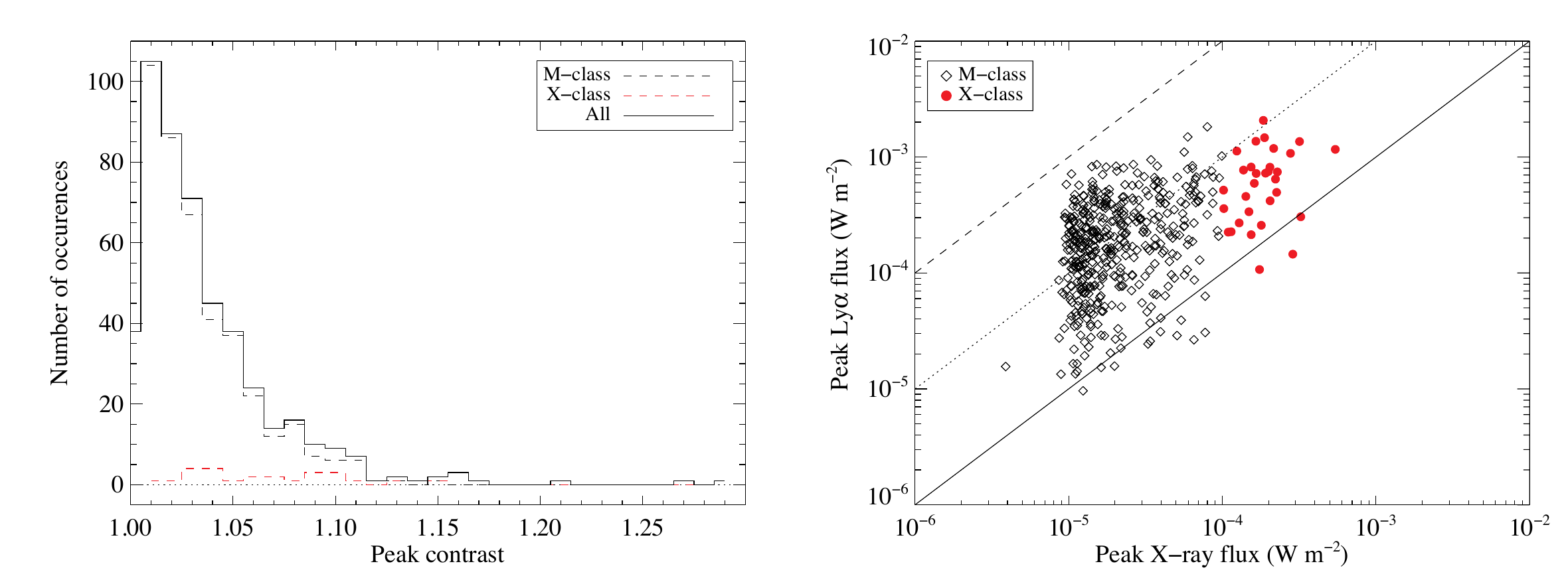}
%\includegraphics[width=0.5\textwidth]{plot_peak_lya_contrast_helio.eps}
%\caption{Scatter plot of the peak flux in \lya\ versus peak flux at 1--8\AA. Black diamonds represent M-class flares, while red solid circles represent X-class flares.}
\caption{Left panel: Histogram of peak contrast of \lya\ emission for all flares in this study. The dashed black curve represents M-class flares while the dashed red curve represents X-classes. The solid black line is the total. Right panel: Scatter plot of the peak excess flux in \lya\ versus peak flux at 1--8\AA. Black diamonds represent M-class flares, while red solid circles represent X-class flares. The solid, dotted, and dashed lines represent the 1:1, 10:1, and 100:1 ratios, respectively.}
%Scatter plot of peak \lya\ emission as a function of heliocentric angle (grey triangles). The black solid circles with error bars denote the mean and 1$\sigma$ distribution at each integer value of angle.
\label{fig:lya_cont}
\end{figure}

\begin{figure}[!h]
\includegraphics[width=\textwidth]{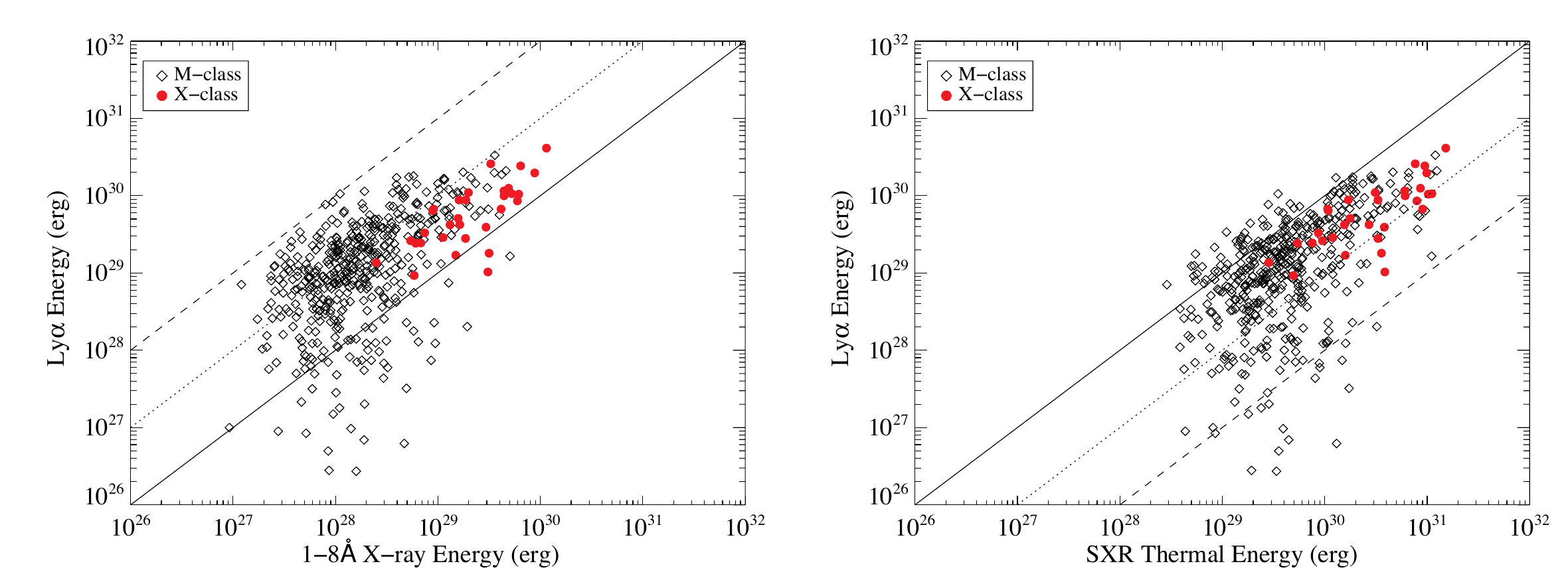}
\caption{Left panel: Scatter plot of total energy radiated by \lya\ against that radiated in the GOES 1--8\AA\ passband. Right panel: Scatter plot of total energy radiated by \lya\ against that radiated by the total SXR emitting plasma. Black diamonds represent M-class flares in both panels, while red solid circles represent X-class flares. The solid, dotted, and dashed lines represent the 1:1, 10:1, and 100:1 ratios, respectively.}
\label{fig:xray_lya_energy}
\end{figure}

Solar flares are known to vary by over two orders of magnitude in the X-ray portion of the spectrum relative to the background emission during the most extreme events \citep{ryan12}, while the corresponding increase in \lya\ irradiance can be marginal due to the much higher background at this wavelength. The left hand panel of Figure~\ref{fig:lya_cont} shows a histogram of the peak \lya\ flux for all flares in this study relative to their pre-flare background. This shows that 95\% of flares exhibit a $\lesssim$10\% enhancement in \lya, with a maximum contrast of $\sim$30\%. These variations are comparable to or greater than changes brought about during active region evolution, albeit on much shorter time scales \citep{wood00}. The right hand panel shows how this peak \lya\ flux varies with the equivalent peak X-ray flux (essentially, GOES class). For M-class flares ($\sim$10$^{-5}$~W~m$^{-2}$), peak \lya\ flux is on average a factor of 10 more intense that that of the X-rays, and as much as a factor of 100 for some events. For X-class flares, the \lya\ flux averages about three times that of X-ray flux, with a maximum of a factor of 10. This is in agreement with \cite{kret15} who found a similar scaling relationship between \lya\ and X-ray fluence for around 100 events.

Previous reports of \lya\ enhancements during flares have often focused on individual or small numbers of events. \cite{wood04} reported a 20\% increase during an X28 flare observed by SORCE/SOLSTICE, while \cite{mill14} reported an 8\% increase during an X2.2 flare observed by SDO/EVE. \cite{kret13} and \cite{raul13} both reported enhancements of $<$1\% using data from PROBA2/LYRA. While the M2 flare presented by \cite{kret13} was not included in this current study, it was observed by GOES-14/EUVS-E which measured a $\sim$3\% increase, suggesting that LYRA may have been underestimating flare-related enhancements to the solar irradiance. This study now presents a comprehensive overview of the variability of solar \lya\ irradiance due to major solar flares through a systematic analysis of almost 500 events.
%While not likely to be the direct cause of this discrepancy, it is worth remembering that PROBA2 is in low-earth orbit and therefore continuously subject to geocoronal absorption, whereas GOES is only affected for a few hours per orbit as certain times of the year.

\subsection{Flare Energetics}
\label{sec:flare_energy}

Both \lya\ and SXR are known drivers of changes in the ionosphere. Therefore knowing the proportional amounts of energy that are injected into the terrestrial atmosphere during flares is important for assessing their relative effects. \cite{mill12} showed that twice as much energy was radiated by \lya\ compared to X-rays during a single X-class flare that occurred close to disk center, while \cite{kret15} found that the \lya\ fluence is around an order of magnitude higher than that of SXR for around 100 flares. The left hand panel of Figure~\ref{fig:xray_lya_energy} shows the total \lya\ energy radiated relative to the total X-ray energy between the GOES start and end times of each event. Similar to that found for the peak fluxes presented in Section~\ref{sec:flare_cont}, and in agreement with \cite{kret15}, approximately 1--100 times more energy was radiated via the \lya\ line than the 1--8\AA\ channel for most M-class flares. X-class flares were only up to 10 times more intense in \lya\ than in X-rays. Some events showed very weak \lya\ energies ($\lesssim$10$^{28}$~erg) as they might not have exhibited an appreciable enhancement above the background due to their proximity to the solar limb, either because of opacity effects along the line of sight, or the footpoints or ribbons being foreshortened or occulted by the solar disk (see Section~\ref{sec:clv}). \cite{kret15} attributed the weaker events in his correlation to possible geocoronal absorption, which is not the case as those events affected were removed from this analysis.

The right-hand panel of Figure~\ref{fig:xray_lya_energy} shows the same \lya\ energies relative to the total thermal energy radiated from the flaring coronal loops as calculated in Section~\ref{sec:energetics}. In this case, the \lya\ energies are comparable to, or around a factor of 10 less, than the total thermal energy. The most reliable measurement to date of a flare's `true' energy stems from observations of flares in the Total Solar Irradiance (TSI) by \cite{wood06} using SORCE/Total Irradiance Monitor (TIM; \citealt{kopp05}). The authors state that the total flare energy is approximately 105 times that of the 1--8\AA\ energy, although this was only measured for four events. \cite{emsl12} have summarized these results, finding that the total instantaneous thermal energy accounted for about one fifth of the detected bolometric energy. The findings presented here imply that \lya\ is indeed a significant radiator of flare energy, as suggested by \cite{mill14}.

\subsection{Center-to-Limb Variation}
\label{sec:clv}

\begin{figure}[!t]
\includegraphics[width=\textwidth]{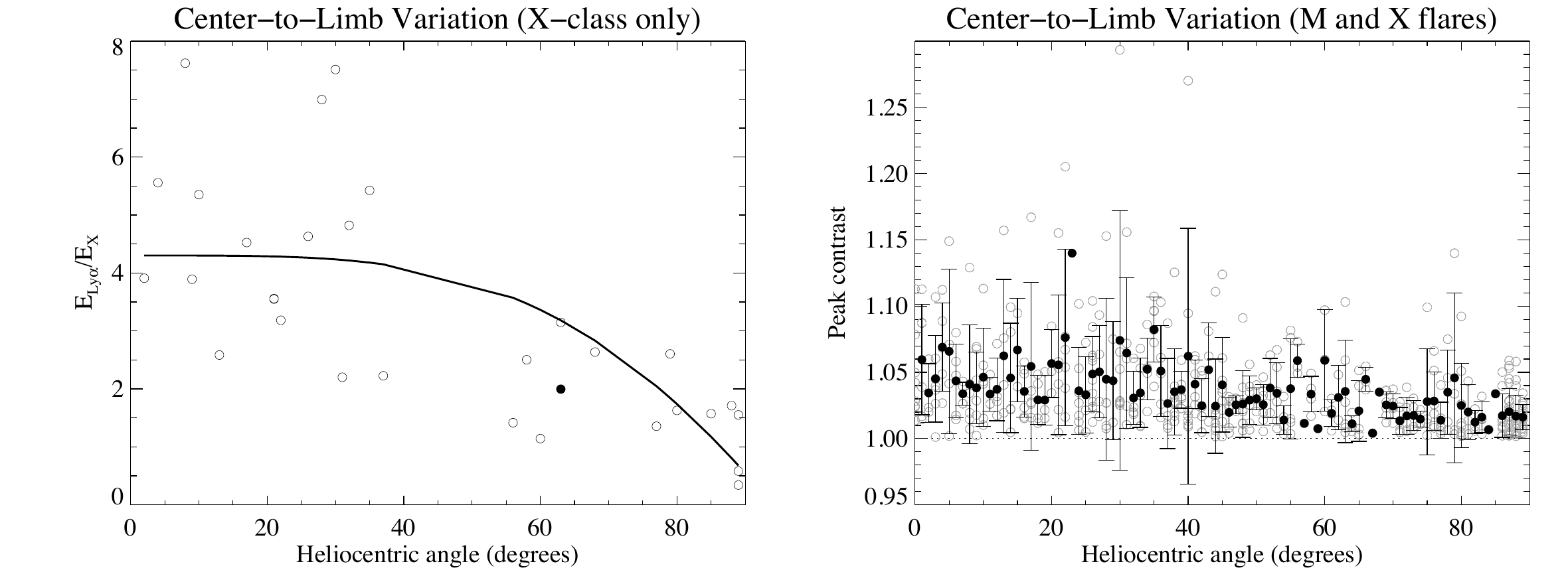}
\caption{Left: Plot of the ratio of energy radiated by \lya\ to that of X-rays, as a function of the heliocentric location of all X-class flares, along with a fit to the data points using Equation~\ref{eqn:clv}. The filled circle denotes the 19 October 2014 flare described in detail in Figure~\ref{fig:maven}. Right: Peak contrast for all flares as a function of heliocentric angle (grey diamonds). The solid black circles denote the mean contrast value at each integer angle, with error bars equivalent to the 1$\sigma$ deviation.}
\label{fig:clv}
\end{figure}

As \lya\ is known to be optically thick \citep[{e.g.,}][]{wood95}, some degree of center-to-limb variation (CLV) is to be expected, given that \lya\ emission from flare ribbons nearer the solar limb will be scattered by the chromospheric column mass along the line of sight. \cite{curd08} showed that this was the case for quiescent \lya\ using spatially-resolved data from SOHO/Solar UV Measurements of Emitted Radiation (SUMER; \citealt{wilh95}). Similarly, \cite{wood06} showed that a CLV was also applicable to the handful of flares observed in the TSI by SORCE/TIM. By measuring the total energy in the TSI flare excess relative to that of the GOES X-rays (which are optically thin, and therefore not attenuated by the solar atmosphere), for each event, the ratio, R, as a function of heliocentric angle was fit with the \textit{ad hoc} quadratic expression
\begin{equation}
R = R_C\left(k+2(1-k)\left(\mu-\frac{\mu^2}{2}\right)\right)\ ,
\label{eqn:clv}
\end{equation}
where $R_C$ is the TSI/X-ray ratio at disk center, $k$ is the limb variation relative to the center, and $\mu$=cos($\theta$) (see \citealt{brek94} for the original empirical derivation). \cite{wood06} found $k=0.11$. The same technique was applied here to \lya\ flares to allow a comparison with previous works. The left-hand panel of Figure~\ref{fig:clv} shows $E_{Ly{\alpha}}/E_{X}$ as a function of heliocentric angle for X-class flares only, with fits to the data using Equation~\ref{eqn:clv} overplotted. The resulting values of $R_C$ and $k$ were 4.33 and 0.12, respectively, in excellent agreement with \cite{wood06}. The right hand panel of Figure~\ref{fig:clv} shows the corresponding peak contrast values of \lya\ for all flares also as a function of heliocentric angles (grey diamonds) with the mean values overplotted as black solid circles with 1$\sigma$ error bars for each integer value of angle. A similar `tailing-off' can be seen towards the limb.
%M-class flares were not included in this anaylsis as the scatter was too great to derive any meaningful conclusions.

\begin{figure}[!h]
\begin{center}
\includegraphics[width=\textwidth]{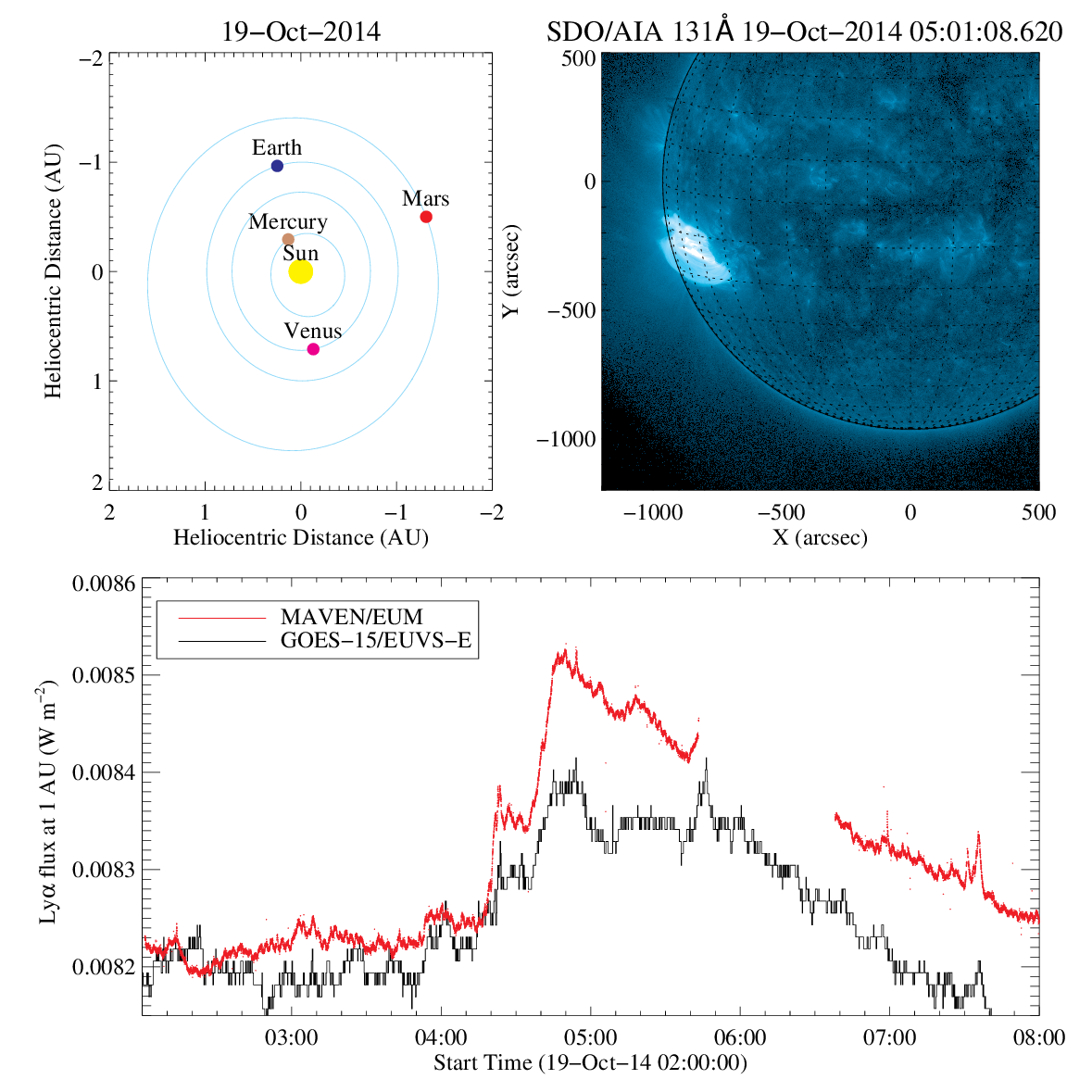}
\caption{Top left panel: Top-down view of the inner solar system on 19 October 2014. Top right panel: SDO/AIA 131\AA\ image of the X-class flare that occurred on that day. From these two panels it can be seen that the flare occurred on the eastern limb as viewed from Earth, but was close to disk center as viewed from Mars. Bottom panel: time profiles of the \lya\ flux at 1~AU as measured by GOES-15/EUVS-E (black curve) and MAVEN/EUM (red curve) after correcting for the Earth-Mars distance and light travel time.}
\label{fig:maven}
\end{center}
\end{figure}

This result can be confirmed with direct, stereoscopic observations of an X-class limb flare (SOL2014-10-19, S13E57, X1.1; denoted by the filled datapoint in the left hand panel of Figure~\ref{fig:clv}). This was was fortuitously observed in \lya\ by both GOES-15/EUVS-E and the Mars Atmospheric and Volatile Evolution (MAVEN) Extreme Ultraviolet Monitor (MAVEN/EUM; \citealt{epar15}), while Earth and Mars were 88.2$^{\circ}$ apart (see top left panel of Figure~\ref{fig:maven}). MAVEN/EUM also includes a \lya\ photometer that observes full-disk solar irradiance at 1~s cadence to investigate the impact of solar radiation on the Martian atmosphere. (See \citealt{cham18} for an overview of flaring activity during the September 2017 period, including \lya\ observations from MAVEN.) From MAVEN's vantage point, this flare (top right panel of Figure~\ref{fig:maven}) would have appeared close to disk center (S11W31 at Mars). The corresponding lightcurves from both instruments (at 1~AU, after correcting for the Earth-Mars distance and light travel time) are plotted in the bottom panel of Figure~\ref{fig:maven}. The $\sim$45\% increase of flare excess emission observed at Mars relative to that seen at Earth, ($\Delta I_{MAVEN}$/$\Delta I_{GOES}$) intensity is in good agreement with that predicted by our limb-darkening function (Equation~\ref{eqn:clv}). Assuming that the X-ray flux at both vantage points would have been the same, the \lya\ energy at $\sim$30$^{\circ}$ would have been 4.5 times higher than that of the X-rays, while at $\sim$60$^{\circ}$ it would have been 3 times higher. This yields a predicted increase of 66\% compared to our measured value of 45\%. Some discrepancy from calibration uncertainties would be expected, or from a more complete form of limb darkening, which might depend upon source geometry as well as limb distance.
%This flare occurred just before GOES-15 began to peer through the geocorona around 07:00 UT on that day.

\subsection{Frequency Distributions}
\label{sec:freq_dist}

An important property of solar flares (along with other systems exhibiting self-organized criticality) is the slope of the frequency distribution of event occurrence. It has been well established for decades that flare distributions can be well represented with a power law of the form:

\begin{equation}
f=Cx^{-\alpha},
\label{eqn:freq_dist}
\end{equation}

\begin{figure}[!t]
\begin{center}
\includegraphics[width=0.75\textwidth]{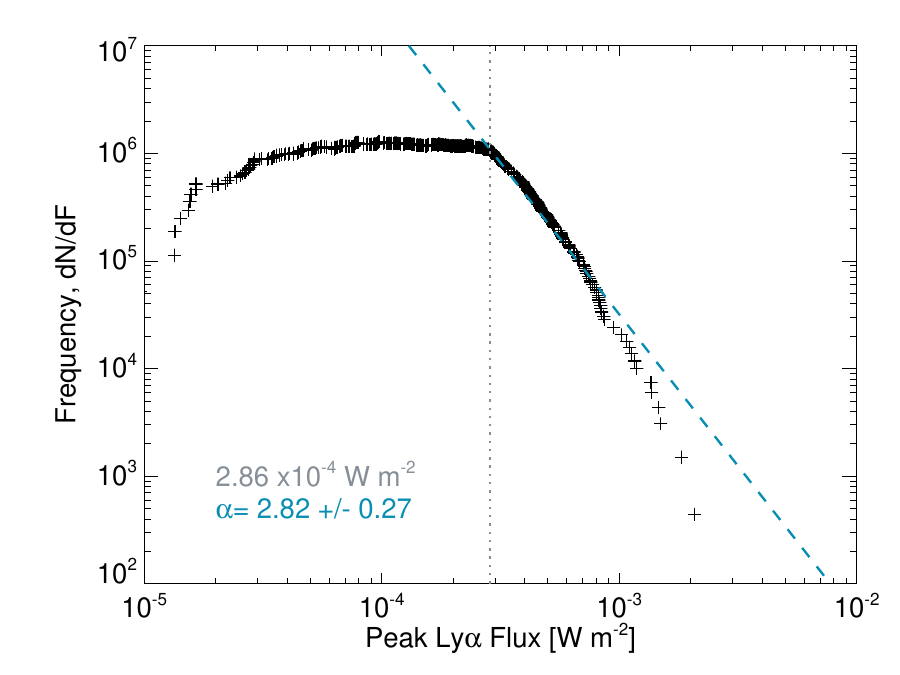}
\caption{Frequency distribution of peak \lya\ fluxes and the resulting power-law fit. This distribution has been found using the technique of \cite{parn00}, with the data points being the empirical frequency function and the dashed line being the fitted power-law above the break (shown as the vertical dotted line).}
\label{fig:freq_dist}
\end{center}
\end{figure}

\noindent
where $f$ is the frequency distribution of a flare parameter, $x$, $\alpha>0$ is the power-law index, and $C$ is a scaling constant. In the case of coronal (e.g. EUV) emission, a value of $\alpha>2$ would imply that, extending the distribution to lower energies, would contain enough energy to heat the solar corona \citep{huds91}. This analysis has since been carried out over a wide range of wavelengths (HXR; \citealt{hann11}, EUV brightenings; \citealt{parn00}, SXR; \citealt{vero02}). Traditionally, the approach has been to bin the data (e.g. peak flux) logarithmically and fit the distribution with a power-law function. \cite{kret15} applied this technique to a sample of $\sim$100 \lya\ flares and derived a value of $\alpha$ between 2.3 and 2.9. However, this method is prone to selection effects, based on the bin sizes \citep[e.g.][]{bai93,ryan16}. Using a Maximum Likelihood method instead removes most of these issues. We adopted the double power-law approach of \citet{parn00}, which not only determines the power-law index, $\alpha$, but also the break in the double power-law, above which the well-sampled distribution can be reliably determined. This flattening of the distribution at lower fluxes is a well known selection effect in which the weaker events are harder to detect and hence undersampled \citep{hann11,asch11}.

The resulting empirical frequency distribution and power-law fit for our peak \lya\ flux is shown in Figure~\ref{fig:freq_dist}. Here an index of $\alpha=2.82 \pm 0.27$ was found for fluxes above $2.86\times10^{-4}~$W~m$^{-2}$. This index lies close to the upper end of the range found by \cite{kret15} and is steeper than those found for the X-ray and other flare observables ($\alpha=1.8$ for HXR bursts; \cite{denn85}, $\alpha=2.1$ for SXR fluxes; \cite{vero02}). Where the empirical frequency distribution matches the power-law line in Figure~\ref{fig:freq_dist} there is a good fit to the data. At higher fluxes we see that there are significantly fewer events than predicted by power law we have found, hinting at a physical effect limiting the event numbers. A similar roll-over has been seen in other flare frequency distribution - for instance, in HXR flare observations from Ulysses \citep{tran09} - but given the infrequency of the largest events, other approaches are often required to better constrain the power law nature of the largest events \citep{schr12}.

%At higher fluxes we see that there are significantly fewer events than predicted by power law we have found, hinting at a physical effect limiting the event numbers; there is no bias against detecting more powerful events. A similar roll-over may have been measured in HXR flare observations from \textit{Ulysses} \citep{tran09}.

\subsection{Ionospheric and Geomagnetic Consequences}
\label{sec:ionosphere}

\begin{figure}[!h]
\begin{center}
\includegraphics[width=\textwidth]{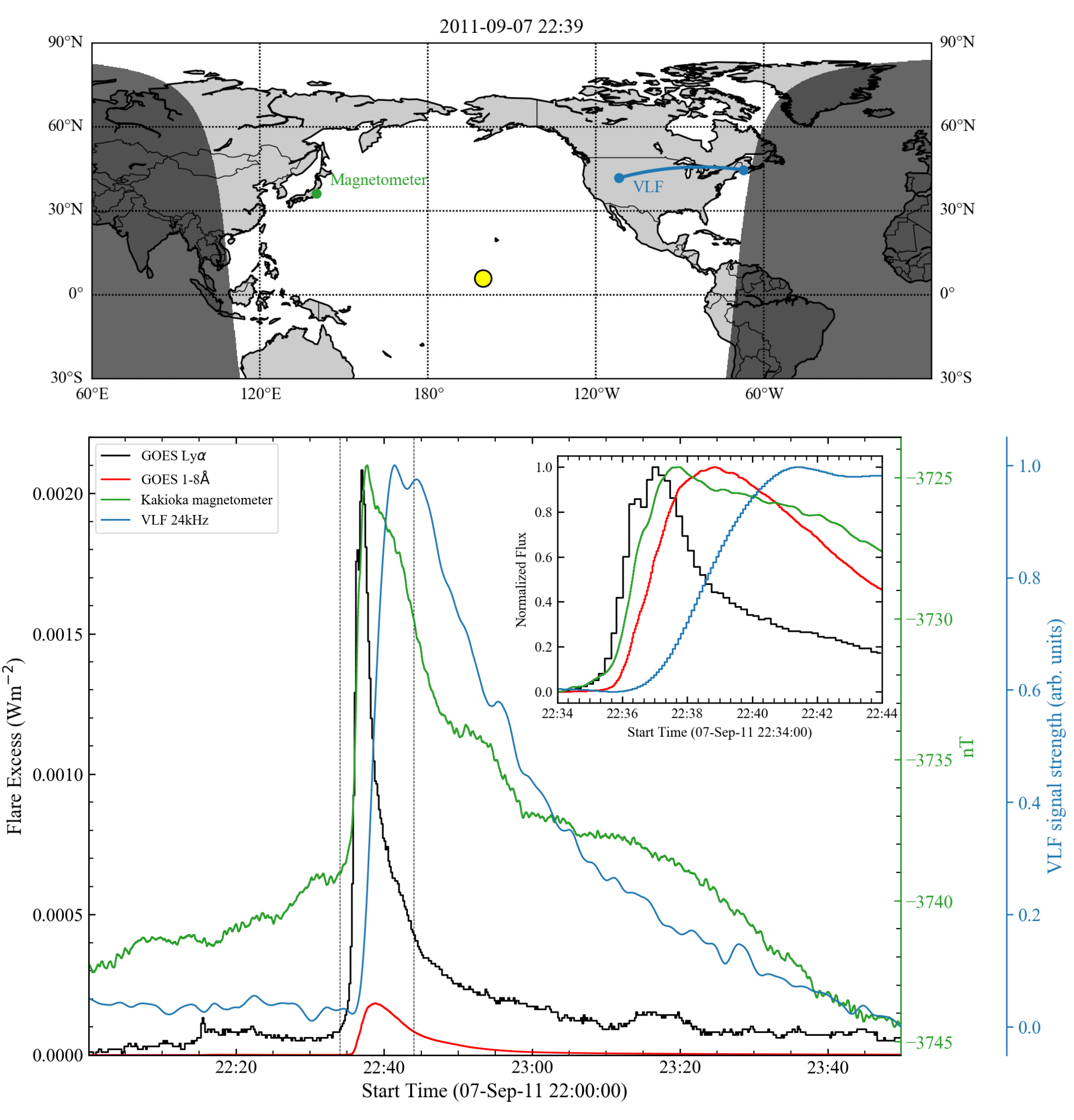}
\caption{The top panel illustrates the locations of the instruments used for the detection of the ionospheric response to the X-class flare on 7 September 2011 (SOL2011-09-07). Both the magnetometer and the VLF path between Maine and Utah were located on the Sun-illuminated portion of the Earth at the time of the flare. The bottom panel shows the time-series of the ionospheric responses of the Sfe recorded in the magnetometer data (green), the VLF amplitude (blue), with flare excess fluxes in \lya\ (black) and 1-8~\AA\ SXR (red). The inset shows all profiles during the rise and peak of the flare (denoted by the two vertical dotted lines in the main panel) normalized to their respective maxima.}
%The solid blue line shows the time derivative of the X-ray signal, a good proxy for the impulsive-phase non-thermal emissions. Note that all time series are relative to their maxima.}
\label{fig:sfe_lya}
\end{center}
\end{figure}

Ionization resulting from solar UV and X-ray emission dominates the Earth's dayside ionosphere. Solar flares provide sudden changes in this ionizing radiation, and \lya\ carries a substantial fraction of the flare perturbations. While disturbances can occur in all regions of the ionosphere following a flare, the effects on the lowest lying regions, typically the D- and E-regions, are most apparent. SXR flare emission is thought to play the largest role as it can vary by several orders of magnitude during major events, but can also penetrate down to low ionospheric altitudes to significantly increase electron density in the D-region. This response can be probed through the observation of Very Low Frequency (VLF 3-30kHz) radio signal amplitude and phase. \lya\ should also have a significant effect in these ionospheric regions, although there has been no comprehensive and systematic monitoring of \lya\ at high time cadence as are now available from GOES. This will help to characterize the ionospheric response to enhanced \lya\ flaring emission.
%correlations with Sfe behavior in order to identify the

The major impulsive flare SOL2011-09-07 was chosen to identify the possible effect of flare-related changes in \lya\ on the ionosphere through observations of enhanced conductivity recorded at the Kakioka magnetometer, Japan and VLF amplitude measurements at a Stanford Sudden Ionospheric Disturbance receiver located in River Heights, Utah. The receiver measures the 24kHz VLF amplitude from the NAA transmitter located in Cutler, Maine (top panel of Figure~\ref{fig:sfe_lya}). The time-series of these observations together with the GOES \lya\ and 1-8\AA\ emission are shown in the bottom panel. This figure shows that the absolute flux increase in \lya\ (black curve) is approximately an order of magnitude greater than that of the SXR (blue curve). The response of the ionosphere can also be clearly identified both in the magnetometer data (E-layer; green) and the VLF data (D-layer; blue). Notably, the inset plot clearly illustrates that the timing of the (normalized) geomagnetic disturbance, known as the ``Solar Flare Effect" (Sfe), closely matches the \lya\ profile, while the SXR emission peaks several minutes later, followed by the VLF amplitude. This indicates that it is not the SXR emission driving the geomagnetic disturbance as observed in the magnetometer data, but the impulsive \lya\ emission.

It is particularly interesting that an impulsive Sfe in the magnetometer (westward) Y-component was observed to closely match the impulsive time scale of \lya, rather than the SXR. This hints at enhanced E-region (90--150~km) current systems perpendicular to the magnetic field due to the ionization of NO by the large energy flux of the \lya\ line, as opposed to ionization in the D-layer which is less electrically conductive due to electron-neutral collisions. This goes against the currently accepted theory that (non-flaring) solar SXR are responsible for ionizing NO \cite[and references therein]{bart88,bart99,sisk90,sisk95,bail02,full04} at these altitudes. We also note that this event produced the largest \lya\ enhancement of all the flares in our study and so may have been a unique case; perhaps NO abundances in the E-layer were significantly higher at the time of this event. There may also have been seasonal or solar cycle effects to consider. A geomagnetic Sfe in fact appeared in conjunction with the first-ever recorded solar flare, SOL1859-09-01 \citep{carr59,stew61} in the form of deflections seen in a ``self-recording magnetograph.''

The D-region response to solar flares, as measured by VLF amplitude observations, often closely matches the SXR flare emission and typically do not show a good correlation with the more impulsive \lya\ \citep[e.g.][]{raul13}. During quiet solar conditions the unperturbed daytime D-region is maintained by \lya\ emission acting on minor constituent NO, however during a flaring event the increased SXR penetrates down to D-region altitudes to dominate ionization of neutral constituents including N$_2$ and O$_2$. This markedly increases electron density in this region large enough to change the propagation conditions of VLF radio waves, and hence the flare response is reflected in the VLF amplitude measurements \citep{thomson_clivard, thomson2011}. The $\sim$3-minute time-delay between the SXR and VLF amplitude, shown in Figure~\ref{fig:sfe_lya}, is characteristic of the D-region, known as the `sluggishness' of the ionosphere \citep{appl53} and signifies the time taken for the D-region photoionization-recombination processes to recover balance after increased ionization \citep[see also][]{zigm07}. This case study however clearly shows that flare \lya\ can have an impact on the ionospheric response (potentially at higher altitudes) in addition to the SXR flux which is typically thought to be the main driver of flare-driven ionization. The availability of high-cadence \lya\ data from GOES-15 (and from future GOES missions) allows for the systematic investigation of the relative timings of the Sfe.

\section{Conclusions}
\label{sec:conc}
This paper presents an overview of over six years of solar flare observations in \lya\ emission (477 events) taken with the E-channel of the EUVS instrument on GOES-15. Prior to this, \lya\ flare observations had been severely limited, with sometimes contradictory results, hampering the understanding of how these events contribute to the broader field of space weather. After removing flaring events that were affected by geocoronal absorption, this work shows that \lya\ is significantly (1--100$\times$) more energetic than the more commonly studied SXR (1--8\AA), which are also a driver of ionospheric disturbances, in agreement with an earlier study of $\sim$100 events by \cite{kret15}. We also show that \lya\ enhancements are comparable to those due to solar rotation, though on timescales of minutes rather than months. However, the \lya\ contribution is less significant for flares that occur close to the solar limb, due to either opacity effects, or foreshortening or occultation of the flare footpoints as seen from Earth. This center-to-limb variation was verified through a simultaneous observations of a limb flare that was also observed by MAVEN/EUM when Earth and Mars were $\sim$90~degrees apart. The frequency distribution of peak \lya\ flux revealed a power law slope of $\alpha$=2.8, implying that the cumulative input from smaller flares may pose a greater energetic input into Earth's atmosphere than the summation of larger events. 

The high time cadence of the GOES-15/EUVS-E data also allows a more comprehensive comparison of the relative ionospheric effects between \lya\ and X-rays. Using magnetometer data from Kakioka it was found that increased conductivity in the ionosphere in response to a major solar flare was highly correlated with with the increase in \lya\ flux, while the corresponding increase in X-rays lagged by several minutes, indicating that the X-rays could not have been the driver of the initial geomagnetic impulse. Our observations suggests that the \lya\ driven Sfe occurs in the E-region, rather than the D-region, and may explain why no such ionospheric response (as characterized by a VLF phase change) was identified by \cite{raul13} during seven solar flares observed by PROBA2/LYRA. The flares that they studied were also of a lower magnitude that those presented here (C- and low M-class), and were therefore unlikely to have produced any enhancement above the solar \lya\ background. Some of these flares also occurred close to the solar limb further diminishing the possibility of them contributing to variations in the solar \lya\ irradiance.

% \cite{raul13} reported no such ionospheric response (as characterized by a VLF phase change) during seven solar flares observed by PROBA2/LYRA.
%, as their observations were analysing the D-region

While the most recent GOES-15 \lya\ data (including those from the period of intense flaring activity in September 2017) have not yet been released at the time of writing, the heliophysics and space weather communities are also anticipating further \lya\ observations from the next generation of GOES satellites, the GOES-R series. These four spacecraft include a dedicated suite of EUV \citep{epar09} and X-ray \citep{cham09} instruments (EUV and X-ray Irradiance Sensors; EXIS), which will provide more advanced coverage of the Sun's output over the next 20 years or more. The \lya\ line will be sampled across five spectral bins, giving a rough estimate of variations in the full-disk line profile.

\lya\ also plays a very important role in trying to understand the physical processes that underpin solar flares themselves. The current study confirms that flare \lya\ emission has a clear impulsive-phase peak. It had been established that \lya\ is a significant radiator of flare energy, but this behavior also means that the data can be used to provide clues as to how the solar chromosphere and transition region respond to flare energy release. Future flare studies could look at the link between \lya\ and LyC \citep{mach18}, or \lya\ and H$\alpha$ \citep{canf81}, for example. The recent launch of Solar Orbiter also included an Extreme Ultraviolet Imager (EUI; \citealt{schu11,roch20}) that contains a \lya\ channel as part of its High Resolution Imager (HRI) suite. EUI will image the Sun in \lya\ at $<$1~s cadence, at 1$''$ resolution at 0.3AU. The currently proposed Japanese Solar-C mission (the follow up to Hinode) is also expected to feature a \lya\ spectrograph \citep{teri11}, as well as the Lyman-alpha Solar Telescope (LST) on the Chinese ASO-S satellite \citep{li2016}. The findings presented here will also assist in the interpretation of results from these future observing platforms.

\acknowledgments
ROM would like to thank Kim Tolbert at NASA/GSFC for her help with making the GOES/EUVS data easily accessible, Janet Machol at NOAA for her advice and assistance, Hamish Reid for several stimulating discussions, and to Edward Cliver for bringing the Kakioka data to our attention; we thank the Kakioka Magnetic Observatory for providing these data (\url{http://www.kakioka-jma.go.jp/obsdata/metadata/en}). ROM would also like to thank Science and Technologies Facilities Council (UK) for the award of an Ernest Rutherford Fellowship (ST/N004981/1), and to the University of Glasgow for a Lord Kelvin Adam Smith award. PCC would like to acknowledge funding from NASA Living with a Star Grant NNX16AE86G titled ``Improving Solar EUV Spectral Irradiance Models with Multi-Vantage Point Observations''. IGH is supported by a Royal Society University Fellowship. LAH is supported by an appointment to the NASA Postdoctoral Program at Goddard Space Flight Center, administered by USRA through a contract with NASA. The GOES and MAVEN data used in this study are freely and publicly available from the instrument websites (GOES/EUVS: \url{https://satdat.ngdc.noaa.gov/sem/goes/data/euvs/GOES_v4/G15/}; GOES/XRS: \url{https://www.ngdc.noaa.gov/stp/satellite/goes/dataaccess.html}; MAVEN/EUM: \url{https://lasp.colorado.edu/lisird/data/mvn_euv_l2_bands/}). The VLF data is available from the Stanford Sudden Ionospheric Disturbance database (\url{http://sid.stanford.edu/database-browser/}). All the authors would like to thank the four anonymous referees for their helpful comments, and editor Mike Hapgood in particular for his support and encouragement. 
%MAVEN data used in this study are available through the NASA Planetary Data System at \url{https://pds.nasa.gov}.

\appendix
\section{Acoustic Oscillations in MAVEN/EUM data}
\label{sec:maven_qpp}

\begin{figure}[!h]
\begin{center}
\includegraphics[width=\textwidth]{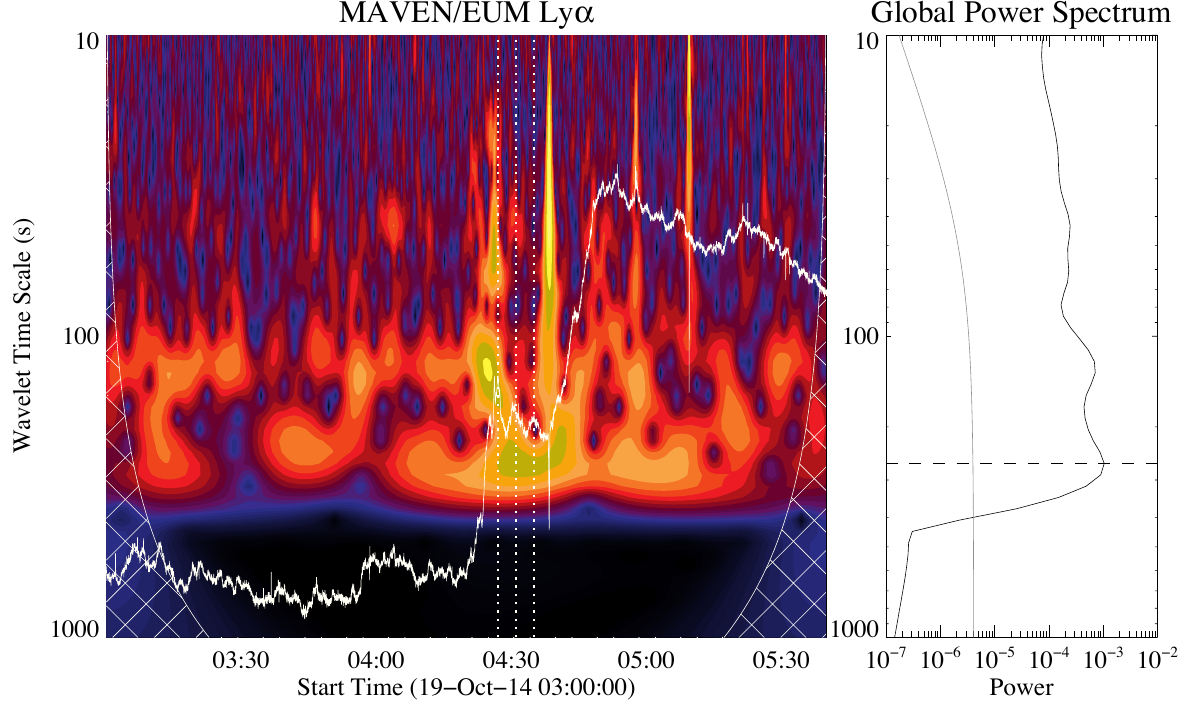}
\caption{Left panel: Wavelet Power Spectrum of \lya\ emission from MAVEN/EUM during the 19 October 2014 flare. The raw lightcurve is overplotted in white. Right panel: Global Power Spectrum integrated over time. Horizontal dashed line denotes a period of 4.4 minutes, while the grey curve denotes the 99\% significance level.}
\label{fig:maven_qpp}
\end{center}
\end{figure}

As mentioned in Section~\ref{sec:intro}, \cite{mill17} detected 3-minute oscillations in full-disk \lya\ emission from GOES-15 during the 15 February 2011 (SOL2011-02-15) X-class flare. This was interpreted as the generation or enhancement of acoustic waves in the chromosphere in response to the impulsive release of energy during the initial stages of the flare. The oscillation was found to be independent of rate of heating due to nonthermal electrons as determined from HXR observations. To date, this is the only reported case of acoustic oscillations in full-disk EUV irradiance data. However, for the X-class flare described in Section~\ref{sec:clv} (SOL2014-10-19), the standard wavelet analysis of \cite{torr98} was applied to the \lya\ lightcurve from MAVEN/EUM. Figure~\ref{fig:maven_qpp} shows that the wavelet power around the onset of the flare is also enhanced at a period of 4.4 minutes, indicative of flare-induced acoustic waves. This deviates slightly from the value found by \cite{mill17} but \cite{jess13} shows that the inclination of the magnetic field can affect the periodicity of propagating disturbances in the chromosphere.

%\bibliography{ms_arxiv.bib}

%% ------------------------------------------------------------------------ %%
%% References and Citations

%%%%%%%%%%%%%%%%%%%%%%%%%%%%%%%%%%%%%%%%%%%%%%%
% BibTeX is preferred:
%
% \bibliography{<name of your .bib file>}

\begin{thebibliography}{}

\bibitem [\protect \citeauthoryear {%
Appleton%
}{%
Appleton%
}{%
{\protect \APACyear {1953}}%
}]{%
appl53}
\APACinsertmetastar {%
appl53}%
\begin{APACrefauthors}%
Appleton, E\BPBI V.%
\end{APACrefauthors}%
\unskip\
\newblock
\APACrefYearMonthDay{1953}{}{}.
\newblock
{\BBOQ}\APACrefatitle {A note on the ÒsluggishnessÓ of the ionosphere} {A
  note on the ÒsluggishnessÓ of the ionosphere}.{\BBCQ}
\newblock
\APACjournalVolNumPages{Journal of Atmospheric and Terrestrial
  Physics}{3}{5}{282--284}.
\PrintBackRefs{\CurrentBib}

\bibitem [\protect \citeauthoryear {%
{Aschwanden}%
}{%
{Aschwanden}%
}{%
{\protect \APACyear {2011}}%
}]{%
asch11}
\APACinsertmetastar {%
asch11}%
\begin{APACrefauthors}%
{Aschwanden}, M\BPBI J.%
\end{APACrefauthors}%
\unskip\
\newblock
\APACrefYearMonthDay{2011}{{\APACmonth{12}}}{}.
\newblock
{\BBOQ}\APACrefatitle {{The State of Self-organized Criticality of the Sun
  During the Last Three Solar Cycles. I. Observations}} {{The State of
  Self-organized Criticality of the Sun During the Last Three Solar Cycles. I.
  Observations}}.{\BBCQ}
\newblock
\APACjournalVolNumPages{\solphys}{274}{1-2}{99-117}.
\newblock
\begin{APACrefDOI} \doi{10.1007/s11207-011-9755-0} \end{APACrefDOI}
\PrintBackRefs{\CurrentBib}

\bibitem [\protect \citeauthoryear {%
{Bai}%
}{%
{Bai}%
}{%
{\protect \APACyear {1993}}%
}]{%
bai93}
\APACinsertmetastar {%
bai93}%
\begin{APACrefauthors}%
{Bai}, T.%
\end{APACrefauthors}%
\unskip\
\newblock
\APACrefYearMonthDay{1993}{{\APACmonth{02}}}{}.
\newblock
{\BBOQ}\APACrefatitle {{Variability of the occurrence frequency of solar flares
  as a function of peak hard X-ray rate}} {{Variability of the occurrence
  frequency of solar flares as a function of peak hard X-ray rate}}.{\BBCQ}
\newblock
\APACjournalVolNumPages{\apj}{404}{}{805-809}.
\newblock
\begin{APACrefDOI} \doi{10.1086/172335} \end{APACrefDOI}
\PrintBackRefs{\CurrentBib}

\bibitem [\protect \citeauthoryear {%
{Bailey}%
, {Barth}%
\BCBL {}\ \BBA {} {Solomon}%
}{%
{Bailey}%
\ \protect \BOthers {.}}{%
{\protect \APACyear {2002}}%
}]{%
bail02}
\APACinsertmetastar {%
bail02}%
\begin{APACrefauthors}%
{Bailey}, S\BPBI M.%
, {Barth}, C\BPBI A.%
\BCBL {}\ \BBA {} {Solomon}, S\BPBI C.%
\end{APACrefauthors}%
\unskip\
\newblock
\APACrefYearMonthDay{2002}{Aug}{}.
\newblock
{\BBOQ}\APACrefatitle {{A model of nitric oxide in the lower thermosphere}} {{A
  model of nitric oxide in the lower thermosphere}}.{\BBCQ}
\newblock
\APACjournalVolNumPages{Journal of Geophysical Research (Space
  Physics)}{107}{A8}{1205}.
\newblock
\begin{APACrefDOI} \doi{10.1029/2001JA000258} \end{APACrefDOI}
\PrintBackRefs{\CurrentBib}

\bibitem [\protect \citeauthoryear {%
Baliukin%
, Bertaux%
, Qu{\'e}merais%
, Izmodenov%
\BCBL {}\ \BBA {} Schmidt%
}{%
Baliukin%
\ \protect \BOthers {.}}{%
{\protect \APACyear {2019}}%
}]{%
bali19}
\APACinsertmetastar {%
bali19}%
\begin{APACrefauthors}%
Baliukin, I\BPBI I.%
, Bertaux, J\BPBI L.%
, Qu{\'e}merais, E.%
, Izmodenov, V\BPBI V.%
\BCBL {}\ \BBA {} Schmidt, W.%
\end{APACrefauthors}%
\unskip\
\newblock
\APACrefYearMonthDay{2019}{{\APACmonth{02}}}{}.
\newblock
{\BBOQ}\APACrefatitle {{SWAN/SOHO Lyman- $\alpha$Mapping: The Hydrogen
  Geocorona Extends Well Beyond the Moon}} {{SWAN/SOHO Lyman- $\alpha$Mapping:
  The Hydrogen Geocorona Extends Well Beyond the Moon}}.{\BBCQ}
\newblock
\APACjournalVolNumPages{Journal of Geophysical Research: Space
  Physics}{124}{2}{861--885}.
\PrintBackRefs{\CurrentBib}

\bibitem [\protect \citeauthoryear {%
{Barth}%
, {Bailey}%
\BCBL {}\ \BBA {} {Solomon}%
}{%
{Barth}%
\ \protect \BOthers {.}}{%
{\protect \APACyear {1999}}%
}]{%
bart99}
\APACinsertmetastar {%
bart99}%
\begin{APACrefauthors}%
{Barth}, C\BPBI A.%
, {Bailey}, S\BPBI M.%
\BCBL {}\ \BBA {} {Solomon}, S\BPBI C.%
\end{APACrefauthors}%
\unskip\
\newblock
\APACrefYearMonthDay{1999}{Jan}{}.
\newblock
{\BBOQ}\APACrefatitle {{Solar-terrestrial coupling: Solar soft X-rays and
  thermospheric nitric oxide}} {{Solar-terrestrial coupling: Solar soft X-rays
  and thermospheric nitric oxide}}.{\BBCQ}
\newblock
\APACjournalVolNumPages{\grl}{26}{9}{1251-1254}.
\newblock
\begin{APACrefDOI} \doi{10.1029/1999GL900237} \end{APACrefDOI}
\PrintBackRefs{\CurrentBib}

\bibitem [\protect \citeauthoryear {%
{Barth}%
, {Tobiska}%
, {Siskind}%
\BCBL {}\ \BBA {} {Cleary}%
}{%
{Barth}%
\ \protect \BOthers {.}}{%
{\protect \APACyear {1988}}%
}]{%
bart88}
\APACinsertmetastar {%
bart88}%
\begin{APACrefauthors}%
{Barth}, C\BPBI A.%
, {Tobiska}, W\BPBI K.%
, {Siskind}, D\BPBI E.%
\BCBL {}\ \BBA {} {Cleary}, D\BPBI D.%
\end{APACrefauthors}%
\unskip\
\newblock
\APACrefYearMonthDay{1988}{Jan}{}.
\newblock
{\BBOQ}\APACrefatitle {{Solar-terrestrial coupling: Low-latitude thermospheric
  nitric oxide}} {{Solar-terrestrial coupling: Low-latitude thermospheric
  nitric oxide}}.{\BBCQ}
\newblock
\APACjournalVolNumPages{\grl}{15}{1}{92-94}.
\newblock
\begin{APACrefDOI} \doi{10.1029/GL015i001p00092} \end{APACrefDOI}
\PrintBackRefs{\CurrentBib}

\bibitem [\protect \citeauthoryear {%
Brekke%
\ \BBA {} Kjeldseth-Moe%
}{%
Brekke%
\ \BBA {} Kjeldseth-Moe%
}{%
{\protect \APACyear {1994}}%
}]{%
brek94}
\APACinsertmetastar {%
brek94}%
\begin{APACrefauthors}%
Brekke, P.%
\BCBT {}\ \BBA {} Kjeldseth-Moe, O.%
\end{APACrefauthors}%
\unskip\
\newblock
\APACrefYearMonthDay{1994}{{\APACmonth{03}}}{}.
\newblock
{\BBOQ}\APACrefatitle {{The solar UV continuum 1440{\textendash}1680 {\AA} and
  its center-to-limb variation}} {{The solar UV continuum 1440{\textendash}1680
  {\AA} and its center-to-limb variation}}.{\BBCQ}
\newblock
\APACjournalVolNumPages{Solar Physics}{150}{1-2}{19--47}.
\PrintBackRefs{\CurrentBib}

\bibitem [\protect \citeauthoryear {%
{Brekke}%
, {Rottman}%
, {Fontenla}%
\BCBL {}\ \BBA {} {Judge}%
}{%
{Brekke}%
\ \protect \BOthers {.}}{%
{\protect \APACyear {1996}}%
}]{%
brek96}
\APACinsertmetastar {%
brek96}%
\begin{APACrefauthors}%
{Brekke}, P.%
, {Rottman}, G\BPBI J.%
, {Fontenla}, J.%
\BCBL {}\ \BBA {} {Judge}, P\BPBI G.%
\end{APACrefauthors}%
\unskip\
\newblock
\APACrefYearMonthDay{1996}{{\APACmonth{09}}}{}.
\newblock
{\BBOQ}\APACrefatitle {{The Ultraviolet Spectrum of a 3B Class Flare Observed
  with SOLSTICE}} {{The Ultraviolet Spectrum of a 3B Class Flare Observed with
  SOLSTICE}}.{\BBCQ}
\newblock
\APACjournalVolNumPages{\apj}{468}{}{418}.
\newblock
\begin{APACrefDOI} \doi{10.1086/177701} \end{APACrefDOI}
\PrintBackRefs{\CurrentBib}

\bibitem [\protect \citeauthoryear {%
Canfield%
, Puetter%
\BCBL {}\ \BBA {} Ricchiazzi%
}{%
Canfield%
\ \protect \BOthers {.}}{%
{\protect \APACyear {1981}}%
}]{%
canf81}
\APACinsertmetastar {%
canf81}%
\begin{APACrefauthors}%
Canfield, R\BPBI C.%
, Puetter, R\BPBI C.%
\BCBL {}\ \BBA {} Ricchiazzi, P\BPBI J.%
\end{APACrefauthors}%
\unskip\
\newblock
\APACrefYearMonthDay{1981}{{\APACmonth{10}}}{}.
\newblock
{\BBOQ}\APACrefatitle {{The Lyman-alpha/H-alpha ratio in solar flares and
  quasars}} {{The Lyman-alpha/H-alpha ratio in solar flares and
  quasars}}.{\BBCQ}
\newblock
\APACjournalVolNumPages{Astrophysical Journal}{249}{}{383--389}.
\PrintBackRefs{\CurrentBib}

\bibitem [\protect \citeauthoryear {%
{Canfield}%
\ \BBA {} {van Hoosier}%
}{%
{Canfield}%
\ \BBA {} {van Hoosier}%
}{%
{\protect \APACyear {1980}}%
}]{%
canf80}
\APACinsertmetastar {%
canf80}%
\begin{APACrefauthors}%
{Canfield}, R\BPBI C.%
\BCBT {}\ \BBA {} {van Hoosier}, M\BPBI E.%
\end{APACrefauthors}%
\unskip\
\newblock
\APACrefYearMonthDay{1980}{{\APACmonth{09}}}{}.
\newblock
{\BBOQ}\APACrefatitle {{Observed L-alpha profiles for two solar flares - 14:12
  UT 15 June, 1973 and 23:16 UT 21 January, 1974}} {{Observed L-alpha profiles
  for two solar flares - 14:12 UT 15 June, 1973 and 23:16 UT 21 January,
  1974}}.{\BBCQ}
\newblock
\APACjournalVolNumPages{\solphys}{67}{}{339-350}.
\newblock
\begin{APACrefDOI} \doi{10.1007/BF00149811} \end{APACrefDOI}
\PrintBackRefs{\CurrentBib}

\bibitem [\protect \citeauthoryear {%
{Carrington}%
}{%
{Carrington}%
}{%
{\protect \APACyear {1859}}%
}]{%
carr59}
\APACinsertmetastar {%
carr59}%
\begin{APACrefauthors}%
{Carrington}, R\BPBI C.%
\end{APACrefauthors}%
\unskip\
\newblock
\APACrefYearMonthDay{1859}{{\APACmonth{11}}}{}.
\newblock
{\BBOQ}\APACrefatitle {{Description of a Singular Appearance seen in the Sun on
  September 1, 1859 }} {{Description of a Singular Appearance seen in the Sun
  on September 1, 1859 }}.{\BBCQ}
\newblock
\APACjournalVolNumPages{Mon. Not. R. Astr. Soc.}{20}{}{13-16}.
\PrintBackRefs{\CurrentBib}

\bibitem [\protect \citeauthoryear {%
Chamberlin%
\ \protect \BOthers {.}}{%
Chamberlin%
\ \protect \BOthers {.}}{%
{\protect \APACyear {2018}}%
}]{%
cham18}
\APACinsertmetastar {%
cham18}%
\begin{APACrefauthors}%
Chamberlin, P\BPBI C.%
, Woods, T\BPBI N.%
, Didkovsky, L.%
, Eparvier, F\BPBI G.%
, Jones, A\BPBI R.%
, Machol, J\BPBI L.%
\BDBL {}Woodraska, D\BPBI L.%
\end{APACrefauthors}%
\unskip\
\newblock
\APACrefYearMonthDay{2018}{{\APACmonth{10}}}{}.
\newblock
{\BBOQ}\APACrefatitle {{Solar Ultraviolet Irradiance Observations of the Solar
  Flares During the Intense September 2017 Storm Period}} {{Solar Ultraviolet
  Irradiance Observations of the Solar Flares During the Intense September 2017
  Storm Period}}.{\BBCQ}
\newblock
\APACjournalVolNumPages{Space Weather}{291}{6}{1665}.
\PrintBackRefs{\CurrentBib}

\bibitem [\protect \citeauthoryear {%
{Chamberlin}%
, {Woods}%
, {Eparvier}%
\BCBL {}\ \BBA {} {Jones}%
}{%
{Chamberlin}%
\ \protect \BOthers {.}}{%
{\protect \APACyear {2009}}%
}]{%
cham09}
\APACinsertmetastar {%
cham09}%
\begin{APACrefauthors}%
{Chamberlin}, P\BPBI C.%
, {Woods}, T\BPBI N.%
, {Eparvier}, F\BPBI G.%
\BCBL {}\ \BBA {} {Jones}, A\BPBI R.%
\end{APACrefauthors}%
\unskip\
\newblock
\APACrefYearMonthDay{2009}{}{}.
\newblock
{\BBOQ}\APACrefatitle {{Next generation x-ray sensor (XRS) for the NOAA GOES-R
  satellite series}} {{Next generation x-ray sensor (XRS) for the NOAA GOES-R
  satellite series}}.{\BBCQ}
\newblock
\BIn{} \APACrefbtitle {Solar Physics and Space Weather Instrumentation III.
  Edited by Fineschi, Silvano; Fennelly, Judy A. Proceedings of the SPIE,
  Volume 7438, article id. 743802, 10 pp. (2009).} {Solar physics and space
  weather instrumentation iii. edited by fineschi, silvano; fennelly, judy a.
  proceedings of the spie, volume 7438, article id. 743802, 10 pp. (2009).}\
  (\BVOL\ 7438, \BPG~743802).
\newblock
\begin{APACrefDOI} \doi{10.1117/12.826807} \end{APACrefDOI}
\PrintBackRefs{\CurrentBib}

\bibitem [\protect \citeauthoryear {%
{Christian}%
\ \protect \BOthers {.}}{%
{Christian}%
\ \protect \BOthers {.}}{%
{\protect \APACyear {2003}}%
}]{%
chri03}
\APACinsertmetastar {%
chri03}%
\begin{APACrefauthors}%
{Christian}, D\BPBI J.%
, {Mathioudakis}, M.%
, {Jevremovi{\'c}}, D.%
, {Dupuis}, J.%
, {Vennes}, S.%
\BCBL {}\ \BBA {} {Kawka}, A.%
\end{APACrefauthors}%
\unskip\
\newblock
\APACrefYearMonthDay{2003}{Aug}{}.
\newblock
{\BBOQ}\APACrefatitle {{The Extreme-Ultraviolet Continuum of a Strong Stellar
  Flare}} {{The Extreme-Ultraviolet Continuum of a Strong Stellar
  Flare}}.{\BBCQ}
\newblock
\APACjournalVolNumPages{\apjl}{593}{2}{L105-L108}.
\newblock
\begin{APACrefDOI} \doi{10.1086/378217} \end{APACrefDOI}
\PrintBackRefs{\CurrentBib}

\bibitem [\protect \citeauthoryear {%
{Chubb}%
, {Friedman}%
, {Kreplin}%
\BCBL {}\ \BBA {} {Kupperian}%
}{%
{Chubb}%
\ \protect \BOthers {.}}{%
{\protect \APACyear {1957}}%
}]{%
chub57}
\APACinsertmetastar {%
chub57}%
\begin{APACrefauthors}%
{Chubb}, T\BPBI A.%
, {Friedman}, H.%
, {Kreplin}, R\BPBI W.%
\BCBL {}\ \BBA {} {Kupperian}, J., J.~E.%
\end{APACrefauthors}%
\unskip\
\newblock
\APACrefYearMonthDay{1957}{Sep}{}.
\newblock
{\BBOQ}\APACrefatitle {{Lyman Alpha and X-Ray Emissions during a Small Solar
  Flare}} {{Lyman Alpha and X-Ray Emissions during a Small Solar
  Flare}}.{\BBCQ}
\newblock
\APACjournalVolNumPages{\jgr}{62}{3}{389-398}.
\newblock
\begin{APACrefDOI} \doi{10.1029/JZ062i003p00389} \end{APACrefDOI}
\PrintBackRefs{\CurrentBib}

\bibitem [\protect \citeauthoryear {%
{Crotser}%
, {Woods}%
, {Eparvier}%
, {Triplett}%
\BCBL {}\ \BBA {} {Woodraska}%
}{%
{Crotser}%
\ \protect \BOthers {.}}{%
{\protect \APACyear {2007}}%
}]{%
crot07}
\APACinsertmetastar {%
crot07}%
\begin{APACrefauthors}%
{Crotser}, D\BPBI A.%
, {Woods}, T\BPBI N.%
, {Eparvier}, F\BPBI G.%
, {Triplett}, M\BPBI A.%
\BCBL {}\ \BBA {} {Woodraska}, D\BPBI L.%
\end{APACrefauthors}%
\unskip\
\newblock
\APACrefYearMonthDay{2007}{}{}.
\newblock
{\BBOQ}\APACrefatitle {{SDO-EVE EUV spectrograph optical design and
  performance}} {{SDO-EVE EUV spectrograph optical design and
  performance}}.{\BBCQ}
\newblock
\BIn{} \APACrefbtitle {\procspie} {\procspie}\ (\BVOL\ 6689, \BPG~66890M).
\newblock
\begin{APACrefDOI} \doi{10.1117/12.732592} \end{APACrefDOI}
\PrintBackRefs{\CurrentBib}

\bibitem [\protect \citeauthoryear {%
{Curdt}%
, {Tian}%
, {Teriaca}%
, {Sch{\"u}hle}%
\BCBL {}\ \BBA {} {Lemaire}%
}{%
{Curdt}%
\ \protect \BOthers {.}}{%
{\protect \APACyear {2008}}%
}]{%
curd08}
\APACinsertmetastar {%
curd08}%
\begin{APACrefauthors}%
{Curdt}, W.%
, {Tian}, H.%
, {Teriaca}, L.%
, {Sch{\"u}hle}, U.%
\BCBL {}\ \BBA {} {Lemaire}, P.%
\end{APACrefauthors}%
\unskip\
\newblock
\APACrefYearMonthDay{2008}{{\APACmonth{12}}}{}.
\newblock
{\BBOQ}\APACrefatitle {{The Ly-{$\alpha$} profile and center-to-limb variation
  of the quiet Sun}} {{The Ly-{$\alpha$} profile and center-to-limb variation
  of the quiet Sun}}.{\BBCQ}
\newblock
\APACjournalVolNumPages{\aap}{492}{}{L9-L12}.
\newblock
\begin{APACrefDOI} \doi{10.1051/0004-6361:200810868} \end{APACrefDOI}
\PrintBackRefs{\CurrentBib}

\bibitem [\protect \citeauthoryear {%
da Costa%
, Fletcher%
, Labrosse%
\BCBL {}\ \BBA {} Zuccarello%
}{%
da Costa%
\ \protect \BOthers {.}}{%
{\protect \APACyear {2009}}%
}]{%
daco09}
\APACinsertmetastar {%
daco09}%
\begin{APACrefauthors}%
da Costa, F\BPBI R.%
, Fletcher, L.%
, Labrosse, N.%
\BCBL {}\ \BBA {} Zuccarello, F.%
\end{APACrefauthors}%
\unskip\
\newblock
\APACrefYearMonthDay{2009}{{\APACmonth{11}}}{}.
\newblock
{\BBOQ}\APACrefatitle {{Observations of a solar flare and filament eruption in
  Lyman $\alpha$ and X-rays}} {{Observations of a solar flare and filament
  eruption in Lyman $\alpha$ and X-rays}}.{\BBCQ}
\newblock
\APACjournalVolNumPages{Astronomy and Astrophysics}{507}{2}{1005--1014}.
\PrintBackRefs{\CurrentBib}

\bibitem [\protect \citeauthoryear {%
Dennis%
}{%
Dennis%
}{%
{\protect \APACyear {1985}}%
}]{%
denn85}
\APACinsertmetastar {%
denn85}%
\begin{APACrefauthors}%
Dennis, B\BPBI R.%
\end{APACrefauthors}%
\unskip\
\newblock
\APACrefYearMonthDay{1985}{{\APACmonth{10}}}{}.
\newblock
{\BBOQ}\APACrefatitle {{Solar hard X-ray bursts}} {{Solar hard X-ray
  bursts}}.{\BBCQ}
\newblock
\APACjournalVolNumPages{Solar Physics}{100}{1-2}{465--490}.
\PrintBackRefs{\CurrentBib}

\bibitem [\protect \citeauthoryear {%
{Dominique}%
\ \protect \BOthers {.}}{%
{Dominique}%
\ \protect \BOthers {.}}{%
{\protect \APACyear {2013}}%
}]{%
domi13}
\APACinsertmetastar {%
domi13}%
\begin{APACrefauthors}%
{Dominique}, M.%
, {Hochedez}, J\BHBI F.%
, {Schmutz}, W.%
, {Dammasch}, I\BPBI E.%
, {Shapiro}, A\BPBI I.%
, {Kretzschmar}, M.%
\BDBL {}{BenMoussa}, A.%
\end{APACrefauthors}%
\unskip\
\newblock
\APACrefYearMonthDay{2013}{{\APACmonth{08}}}{}.
\newblock
{\BBOQ}\APACrefatitle {{The LYRA Instrument Onboard PROBA2: Description and
  In-Flight Performance}} {{The LYRA Instrument Onboard PROBA2: Description and
  In-Flight Performance}}.{\BBCQ}
\newblock
\APACjournalVolNumPages{\solphys}{286}{}{21-42}.
\newblock
\begin{APACrefDOI} \doi{10.1007/s11207-013-0252-5} \end{APACrefDOI}
\PrintBackRefs{\CurrentBib}

\bibitem [\protect \citeauthoryear {%
{Emslie}%
\ \protect \BOthers {.}}{%
{Emslie}%
\ \protect \BOthers {.}}{%
{\protect \APACyear {2012}}%
}]{%
emsl12}
\APACinsertmetastar {%
emsl12}%
\begin{APACrefauthors}%
{Emslie}, A\BPBI G.%
, {Dennis}, B\BPBI R.%
, {Shih}, A\BPBI Y.%
, {Chamberlin}, P\BPBI C.%
, {Mewaldt}, R\BPBI A.%
, {Moore}, C\BPBI S.%
\BDBL {}{Welsch}, B\BPBI T.%
\end{APACrefauthors}%
\unskip\
\newblock
\APACrefYearMonthDay{2012}{{\APACmonth{11}}}{}.
\newblock
{\BBOQ}\APACrefatitle {{Global Energetics of Thirty-eight Large Solar Eruptive
  Events}} {{Global Energetics of Thirty-eight Large Solar Eruptive
  Events}}.{\BBCQ}
\newblock
\APACjournalVolNumPages{\apj}{759}{}{71}.
\newblock
\begin{APACrefDOI} \doi{10.1088/0004-637X/759/1/71} \end{APACrefDOI}
\PrintBackRefs{\CurrentBib}

\bibitem [\protect \citeauthoryear {%
Eparvier%
, Chamberlin%
, Woods%
\BCBL {}\ \BBA {} Thiemann%
}{%
Eparvier%
\ \protect \BOthers {.}}{%
{\protect \APACyear {2015}}%
}]{%
epar15}
\APACinsertmetastar {%
epar15}%
\begin{APACrefauthors}%
Eparvier, F\BPBI G.%
, Chamberlin, P\BPBI C.%
, Woods, T\BPBI N.%
\BCBL {}\ \BBA {} Thiemann, E\BPBI M\BPBI B.%
\end{APACrefauthors}%
\unskip\
\newblock
\APACrefYearMonthDay{2015}{{\APACmonth{09}}}{}.
\newblock
{\BBOQ}\APACrefatitle {{The Solar Extreme Ultraviolet Monitor for MAVEN}} {{The
  Solar Extreme Ultraviolet Monitor for MAVEN}}.{\BBCQ}
\newblock
\APACjournalVolNumPages{Space Science Reviews}{-1}{}{80}.
\PrintBackRefs{\CurrentBib}

\bibitem [\protect \citeauthoryear {%
{Eparvier}%
\ \protect \BOthers {.}}{%
{Eparvier}%
\ \protect \BOthers {.}}{%
{\protect \APACyear {2009}}%
}]{%
epar09}
\APACinsertmetastar {%
epar09}%
\begin{APACrefauthors}%
{Eparvier}, F\BPBI G.%
, {Crotser}, D.%
, {Jones}, A\BPBI R.%
, {McClintock}, W\BPBI E.%
, {Snow}, M.%
\BCBL {}\ \BBA {} {Woods}, T\BPBI N.%
\end{APACrefauthors}%
\unskip\
\newblock
\APACrefYearMonthDay{2009}{}{}.
\newblock
{\BBOQ}\APACrefatitle {{The Extreme Ultraviolet Sensor (EUVS) for GOES-R}}
  {{The Extreme Ultraviolet Sensor (EUVS) for GOES-R}}.{\BBCQ}
\newblock
\BIn{} \APACrefbtitle {Solar Physics and Space Weather Instrumentation III.
  Edited by Fineschi, Silvano; Fennelly, Judy A. Proceedings of the SPIE,
  Volume 7438, article id. 743804, 8 pp. (2009).} {Solar physics and space
  weather instrumentation iii. edited by fineschi, silvano; fennelly, judy a.
  proceedings of the spie, volume 7438, article id. 743804, 8 pp. (2009).}\
  (\BVOL\ 7438, \BPG~743804).
\newblock
\begin{APACrefDOI} \doi{10.1117/12.826445} \end{APACrefDOI}
\PrintBackRefs{\CurrentBib}

\bibitem [\protect \citeauthoryear {%
Evans%
\ \protect \BOthers {.}}{%
Evans%
\ \protect \BOthers {.}}{%
{\protect \APACyear {2010}}%
}]{%
evan10}
\APACinsertmetastar {%
evan10}%
\begin{APACrefauthors}%
Evans, J\BPBI S.%
, Strickland, D\BPBI J.%
, Woo, W\BPBI K.%
, McMullin, D\BPBI R.%
, Plunkett, S\BPBI P.%
, Viereck, R\BPBI A.%
\BDBL {}Eparvier, F\BPBI G.%
\end{APACrefauthors}%
\unskip\
\newblock
\APACrefYearMonthDay{2010}{{\APACmonth{01}}}{}.
\newblock
{\BBOQ}\APACrefatitle {{Early Observations by the GOES-13 Solar Extreme
  Ultraviolet Sensor (EUVS)}} {{Early Observations by the GOES-13 Solar Extreme
  Ultraviolet Sensor (EUVS)}}.{\BBCQ}
\newblock
\APACjournalVolNumPages{Solar Physics}{262}{1}{71--115}.
\PrintBackRefs{\CurrentBib}

\bibitem [\protect \citeauthoryear {%
{Freeland}%
\ \BBA {} {Handy}%
}{%
{Freeland}%
\ \BBA {} {Handy}%
}{%
{\protect \APACyear {1998}}%
}]{%
free98}
\APACinsertmetastar {%
free98}%
\begin{APACrefauthors}%
{Freeland}, S\BPBI L.%
\BCBT {}\ \BBA {} {Handy}, B\BPBI N.%
\end{APACrefauthors}%
\unskip\
\newblock
\APACrefYearMonthDay{1998}{Oct}{}.
\newblock
{\BBOQ}\APACrefatitle {{Data Analysis with the SolarSoft System}} {{Data
  Analysis with the SolarSoft System}}.{\BBCQ}
\newblock
\APACjournalVolNumPages{\solphys}{182}{2}{497-500}.
\newblock
\begin{APACrefDOI} \doi{10.1023/A:1005038224881} \end{APACrefDOI}
\PrintBackRefs{\CurrentBib}

\bibitem [\protect \citeauthoryear {%
{Fuller-Rowell}%
, {Solomon}%
, {Roble}%
\BCBL {}\ \BBA {} {Viereck}%
}{%
{Fuller-Rowell}%
\ \protect \BOthers {.}}{%
{\protect \APACyear {2004}}%
}]{%
full04}
\APACinsertmetastar {%
full04}%
\begin{APACrefauthors}%
{Fuller-Rowell}, T.%
, {Solomon}, S.%
, {Roble}, R.%
\BCBL {}\ \BBA {} {Viereck}, R.%
\end{APACrefauthors}%
\unskip\
\newblock
\APACrefYearMonthDay{2004}{Jan}{}.
\newblock
{\BBOQ}\APACrefatitle {{Impact of Solar EUV, XUV, and X-Ray Variations on
  Earth's Atmosphere}} {{Impact of Solar EUV, XUV, and X-Ray Variations on
  Earth's Atmosphere}}.{\BBCQ}
\newblock
\BIn{} \APACrefbtitle {Solar Variability and its Effects on Climate.
  Geophysical Monograph 141} {Solar variability and its effects on climate.
  geophysical monograph 141}\ (\BVOL~141, \BPG~341).
\newblock
\begin{APACrefDOI} \doi{10.1029/141GM23} \end{APACrefDOI}
\PrintBackRefs{\CurrentBib}

\bibitem [\protect \citeauthoryear {%
{Garcia}%
}{%
{Garcia}%
}{%
{\protect \APACyear {1994}}%
}]{%
garc94}
\APACinsertmetastar {%
garc94}%
\begin{APACrefauthors}%
{Garcia}, H\BPBI A.%
\end{APACrefauthors}%
\unskip\
\newblock
\APACrefYearMonthDay{1994}{Oct}{}.
\newblock
{\BBOQ}\APACrefatitle {{Temperature and Emission Measure from Goes Soft X-Ray
  Measurements}} {{Temperature and Emission Measure from Goes Soft X-Ray
  Measurements}}.{\BBCQ}
\newblock
\APACjournalVolNumPages{\solphys}{154}{2}{275-308}.
\newblock
\begin{APACrefDOI} \doi{10.1007/BF00681100} \end{APACrefDOI}
\PrintBackRefs{\CurrentBib}

\bibitem [\protect \citeauthoryear {%
{Handy}%
\ \protect \BOthers {.}}{%
{Handy}%
\ \protect \BOthers {.}}{%
{\protect \APACyear {1999}}%
}]{%
hand99}
\APACinsertmetastar {%
hand99}%
\begin{APACrefauthors}%
{Handy}, B\BPBI N.%
, {Acton}, L\BPBI W.%
, {Kankelborg}, C\BPBI C.%
, {Wolfson}, C\BPBI J.%
, {Akin}, D\BPBI J.%
, {Bruner}, M\BPBI E.%
\BDBL {}{Parkinson}, C.%
\end{APACrefauthors}%
\unskip\
\newblock
\APACrefYearMonthDay{1999}{{\APACmonth{07}}}{}.
\newblock
{\BBOQ}\APACrefatitle {{The transition region and coronal explorer}} {{The
  transition region and coronal explorer}}.{\BBCQ}
\newblock
\APACjournalVolNumPages{\solphys}{187}{}{229-260}.
\newblock
\begin{APACrefDOI} \doi{10.1023/A:1005166902804} \end{APACrefDOI}
\PrintBackRefs{\CurrentBib}

\bibitem [\protect \citeauthoryear {%
Hannah%
\ \protect \BOthers {.}}{%
Hannah%
\ \protect \BOthers {.}}{%
{\protect \APACyear {2011}}%
}]{%
hann11}
\APACinsertmetastar {%
hann11}%
\begin{APACrefauthors}%
Hannah, I\BPBI G.%
, Hudson, H\BPBI S.%
, Battaglia, M.%
, Christe, S.%
, Ka{\v s}parov{\'a}, J.%
, Krucker, S.%
\BDBL {}Veronig, A.%
\end{APACrefauthors}%
\unskip\
\newblock
\APACrefYearMonthDay{2011}{{\APACmonth{07}}}{}.
\newblock
{\BBOQ}\APACrefatitle {{Microflares and the Statistics of X-ray Flares}}
  {{Microflares and the Statistics of X-ray Flares}}.{\BBCQ}
\newblock
\APACjournalVolNumPages{Space Science Reviews}{159}{1-4}{263--300}.
\PrintBackRefs{\CurrentBib}

\bibitem [\protect \citeauthoryear {%
{Hanser}%
\ \BBA {} {Sellers}%
}{%
{Hanser}%
\ \BBA {} {Sellers}%
}{%
{\protect \APACyear {1996}}%
}]{%
hans96}
\APACinsertmetastar {%
hans96}%
\begin{APACrefauthors}%
{Hanser}, F\BPBI A.%
\BCBT {}\ \BBA {} {Sellers}, F\BPBI B.%
\end{APACrefauthors}%
\unskip\
\newblock
\APACrefYearMonthDay{1996}{{\APACmonth{10}}}{}.
\newblock
{\BBOQ}\APACrefatitle {{Design and calibration of the GOES-8 solar x-ray
  sensor: the XRS}} {{Design and calibration of the GOES-8 solar x-ray sensor:
  the XRS}}.{\BBCQ}
\newblock
\BIn{} E\BPBI R.~{Washwell}\ (\BED), \APACrefbtitle {GOES-8 and Beyond} {Goes-8
  and beyond}\ (\BVOL\ 2812, \BPG~344-352).
\newblock
\begin{APACrefDOI} \doi{10.1117/12.254082} \end{APACrefDOI}
\PrintBackRefs{\CurrentBib}

\bibitem [\protect \citeauthoryear {%
{Hochedez}%
\ \protect \BOthers {.}}{%
{Hochedez}%
\ \protect \BOthers {.}}{%
{\protect \APACyear {2006}}%
}]{%
hoch06}
\APACinsertmetastar {%
hoch06}%
\begin{APACrefauthors}%
{Hochedez}, J\BHBI F.%
, {Schmutz}, W.%
, {Stockman}, Y.%
, {Sch{\"u}hle}, U.%
, {Benmoussa}, A.%
, {Koller}, S.%
\BDBL {}{Rochus}, P.%
\end{APACrefauthors}%
\unskip\
\newblock
\APACrefYearMonthDay{2006}{}{}.
\newblock
{\BBOQ}\APACrefatitle {{LYRA, a solar UV radiometer on Proba2}} {{LYRA, a solar
  UV radiometer on Proba2}}.{\BBCQ}
\newblock
\APACjournalVolNumPages{Advances in Space Research}{37}{}{303-312}.
\newblock
\begin{APACrefDOI} \doi{10.1016/j.asr.2005.10.041} \end{APACrefDOI}
\PrintBackRefs{\CurrentBib}

\bibitem [\protect \citeauthoryear {%
Hudson%
}{%
Hudson%
}{%
{\protect \APACyear {1991}}%
}]{%
huds91}
\APACinsertmetastar {%
huds91}%
\begin{APACrefauthors}%
Hudson, H\BPBI S.%
\end{APACrefauthors}%
\unskip\
\newblock
\APACrefYearMonthDay{1991}{{\APACmonth{06}}}{}.
\newblock
{\BBOQ}\APACrefatitle {{Solar flares, microflares, nanoflares, and coronal
  heating}} {{Solar flares, microflares, nanoflares, and coronal
  heating}}.{\BBCQ}
\newblock
\APACjournalVolNumPages{Solar Physics}{133}{2}{357--369}.
\PrintBackRefs{\CurrentBib}

\bibitem [\protect \citeauthoryear {%
{Jess}%
, {Reznikova}%
, {Van Doorsselaere}%
, {Keys}%
\BCBL {}\ \BBA {} {Mackay}%
}{%
{Jess}%
\ \protect \BOthers {.}}{%
{\protect \APACyear {2013}}%
}]{%
jess13}
\APACinsertmetastar {%
jess13}%
\begin{APACrefauthors}%
{Jess}, D\BPBI B.%
, {Reznikova}, V\BPBI E.%
, {Van Doorsselaere}, T.%
, {Keys}, P\BPBI H.%
\BCBL {}\ \BBA {} {Mackay}, D\BPBI H.%
\end{APACrefauthors}%
\unskip\
\newblock
\APACrefYearMonthDay{2013}{Dec}{}.
\newblock
{\BBOQ}\APACrefatitle {{The Influence of the Magnetic Field on Running
  Penumbral Waves in the Solar Chromosphere}} {{The Influence of the Magnetic
  Field on Running Penumbral Waves in the Solar Chromosphere}}.{\BBCQ}
\newblock
\APACjournalVolNumPages{\apj}{779}{2}{168}.
\newblock
\begin{APACrefDOI} \doi{10.1088/0004-637X/779/2/168} \end{APACrefDOI}
\PrintBackRefs{\CurrentBib}

\bibitem [\protect \citeauthoryear {%
{Kopp}%
\ \BBA {} {Lawrence}%
}{%
{Kopp}%
\ \BBA {} {Lawrence}%
}{%
{\protect \APACyear {2005}}%
}]{%
kopp05}
\APACinsertmetastar {%
kopp05}%
\begin{APACrefauthors}%
{Kopp}, G.%
\BCBT {}\ \BBA {} {Lawrence}, G.%
\end{APACrefauthors}%
\unskip\
\newblock
\APACrefYearMonthDay{2005}{{\APACmonth{08}}}{}.
\newblock
{\BBOQ}\APACrefatitle {{The Total Irradiance Monitor (TIM): Instrument Design}}
  {{The Total Irradiance Monitor (TIM): Instrument Design}}.{\BBCQ}
\newblock
\APACjournalVolNumPages{\solphys}{230}{}{91-109}.
\newblock
\begin{APACrefDOI} \doi{10.1007/s11207-005-7446-4} \end{APACrefDOI}
\PrintBackRefs{\CurrentBib}

\bibitem [\protect \citeauthoryear {%
{Kretzschmar}%
}{%
{Kretzschmar}%
}{%
{\protect \APACyear {2015}}%
}]{%
kret15}
\APACinsertmetastar {%
kret15}%
\begin{APACrefauthors}%
{Kretzschmar}, M.%
\end{APACrefauthors}%
\unskip\
\newblock
\APACrefYearMonthDay{2015}{Dec}{}.
\newblock
{\BBOQ}\APACrefatitle {{Temperature Dependence of the Flare Fluence Scaling
  Exponent}} {{Temperature Dependence of the Flare Fluence Scaling
  Exponent}}.{\BBCQ}
\newblock
\APACjournalVolNumPages{\solphys}{290}{12}{3593-3609}.
\newblock
\begin{APACrefDOI} \doi{10.1007/s11207-015-0783-z} \end{APACrefDOI}
\PrintBackRefs{\CurrentBib}

\bibitem [\protect \citeauthoryear {%
{Kretzschmar}%
, {Dominique}%
\BCBL {}\ \BBA {} {Dammasch}%
}{%
{Kretzschmar}%
\ \protect \BOthers {.}}{%
{\protect \APACyear {2013}}%
}]{%
kret13}
\APACinsertmetastar {%
kret13}%
\begin{APACrefauthors}%
{Kretzschmar}, M.%
, {Dominique}, M.%
\BCBL {}\ \BBA {} {Dammasch}, I\BPBI E.%
\end{APACrefauthors}%
\unskip\
\newblock
\APACrefYearMonthDay{2013}{{\APACmonth{08}}}{}.
\newblock
{\BBOQ}\APACrefatitle {{Sun-as-a-Star Observation of Flares in Lyman {$\alpha$}
  by the PROBA2/LYRA Radiometer}} {{Sun-as-a-Star Observation of Flares in
  Lyman {$\alpha$} by the PROBA2/LYRA Radiometer}}.{\BBCQ}
\newblock
\APACjournalVolNumPages{\solphys}{286}{}{221-239}.
\newblock
\begin{APACrefDOI} \doi{10.1007/s11207-012-0175-6} \end{APACrefDOI}
\PrintBackRefs{\CurrentBib}

\bibitem [\protect \citeauthoryear {%
Lean%
}{%
Lean%
}{%
{\protect \APACyear {1985}}%
}]{%
lean85}
\APACinsertmetastar {%
lean85}%
\begin{APACrefauthors}%
Lean, J\BPBI L.%
\end{APACrefauthors}%
\unskip\
\newblock
\APACrefYearMonthDay{1985}{}{}.
\newblock
{\BBOQ}\APACrefatitle {{Calculations of Lyman Alpha Absorption in the
  Mesosphere}} {{Calculations of Lyman Alpha Absorption in the
  Mesosphere}}.{\BBCQ}
\newblock
\BIn{} \APACrefbtitle {Atmospheric Ozone} {Atmospheric ozone}\ (\BPGS\
  697--701).
\newblock
\APACaddressPublisher{Dordrecht}{Springer Netherlands}.
\PrintBackRefs{\CurrentBib}

\bibitem [\protect \citeauthoryear {%
{Lean}%
\ \BBA {} {Skumanich}%
}{%
{Lean}%
\ \BBA {} {Skumanich}%
}{%
{\protect \APACyear {1983}}%
}]{%
lean83}
\APACinsertmetastar {%
lean83}%
\begin{APACrefauthors}%
{Lean}, J\BPBI L.%
\BCBT {}\ \BBA {} {Skumanich}, A.%
\end{APACrefauthors}%
\unskip\
\newblock
\APACrefYearMonthDay{1983}{{\APACmonth{07}}}{}.
\newblock
{\BBOQ}\APACrefatitle {{Variability of the Lyman alpha flux with solar
  activity}} {{Variability of the Lyman alpha flux with solar
  activity}}.{\BBCQ}
\newblock
\APACjournalVolNumPages{\jgr}{88}{}{5751-5759}.
\newblock
\begin{APACrefDOI} \doi{10.1029/JA088iA07p05751} \end{APACrefDOI}
\PrintBackRefs{\CurrentBib}

\bibitem [\protect \citeauthoryear {%
Lemen%
\ \protect \BOthers {.}}{%
Lemen%
\ \protect \BOthers {.}}{%
{\protect \APACyear {2011}}%
}]{%
leme11}
\APACinsertmetastar {%
leme11}%
\begin{APACrefauthors}%
Lemen, J\BPBI R.%
, Title, A\BPBI M.%
, Akin, D\BPBI J.%
, Boerner, P\BPBI F.%
, Chou, C.%
, Drake, J\BPBI F.%
\BDBL {}Waltham, N.%
\end{APACrefauthors}%
\unskip\
\newblock
\APACrefYearMonthDay{2011}{{\APACmonth{06}}}{}.
\newblock
{\BBOQ}\APACrefatitle {{The Atmospheric Imaging Assembly (AIA) on the Solar
  Dynamics Observatory (SDO)}} {{The Atmospheric Imaging Assembly (AIA) on the
  Solar Dynamics Observatory (SDO)}}.{\BBCQ}
\newblock
\APACjournalVolNumPages{Solar Physics}{}{}{172}.
\PrintBackRefs{\CurrentBib}

\bibitem [\protect \citeauthoryear {%
Li%
}{%
Li%
}{%
{\protect \APACyear {2016}}%
}]{%
li2016}
\APACinsertmetastar {%
li2016}%
\begin{APACrefauthors}%
Li, H.%
\end{APACrefauthors}%
\unskip\
\newblock
\APACrefYearMonthDay{2016}{}{}.
\newblock
{\BBOQ}\APACrefatitle {{The Lyman-alpha Solar Telescope (LST) for the ASO-S
  mission}} {{The Lyman-alpha Solar Telescope (LST) for the ASO-S
  mission}}.{\BBCQ}
\newblock
\APACjournalVolNumPages{Solar and Stellar Flares and their Effects on
  Planets}{320}{S320}{436--438}.
\PrintBackRefs{\CurrentBib}

\bibitem [\protect \citeauthoryear {%
Lilensten%
\ \protect \BOthers {.}}{%
Lilensten%
\ \protect \BOthers {.}}{%
{\protect \APACyear {2008}}%
}]{%
lile08}
\APACinsertmetastar {%
lile08}%
\begin{APACrefauthors}%
Lilensten, J.%
, Dudok~de Wit, T.%
, Kretzschmar%
, Amblard, P\BPBI O.%
, Moussaoui, S.%
, Aboudarham, J.%
\BCBL {}\ \BBA {} Auch{\`e}re, F.%
\end{APACrefauthors}%
\unskip\
\newblock
\APACrefYearMonthDay{2008}{{\APACmonth{02}}}{}.
\newblock
{\BBOQ}\APACrefatitle {{Review on the solar spectral variability in the EUV for
  space weather purposes}} {{Review on the solar spectral variability in the
  EUV for space weather purposes}}.{\BBCQ}
\newblock
\APACjournalVolNumPages{Annales Geophysicae}{26}{2}{269--279}.
\PrintBackRefs{\CurrentBib}

\bibitem [\protect \citeauthoryear {%
{Linsky}%
, {France}%
\BCBL {}\ \BBA {} {Ayres}%
}{%
{Linsky}%
\ \protect \BOthers {.}}{%
{\protect \APACyear {2013}}%
}]{%
lins13}
\APACinsertmetastar {%
lins13}%
\begin{APACrefauthors}%
{Linsky}, J\BPBI L.%
, {France}, K.%
\BCBL {}\ \BBA {} {Ayres}, T.%
\end{APACrefauthors}%
\unskip\
\newblock
\APACrefYearMonthDay{2013}{{\APACmonth{04}}}{}.
\newblock
{\BBOQ}\APACrefatitle {{Computing Intrinsic LY{$\alpha$} Fluxes of F5 V to M5 V
  Stars}} {{Computing Intrinsic LY{$\alpha$} Fluxes of F5 V to M5 V
  Stars}}.{\BBCQ}
\newblock
\APACjournalVolNumPages{\apj}{766}{}{69}.
\newblock
\begin{APACrefDOI} \doi{10.1088/0004-637X/766/2/69} \end{APACrefDOI}
\PrintBackRefs{\CurrentBib}

\bibitem [\protect \citeauthoryear {%
Machado%
, Milligan%
\BCBL {}\ \BBA {} Simoes%
}{%
Machado%
\ \protect \BOthers {.}}{%
{\protect \APACyear {2018}}%
}]{%
mach18}
\APACinsertmetastar {%
mach18}%
\begin{APACrefauthors}%
Machado, M\BPBI E.%
, Milligan, R\BPBI O.%
\BCBL {}\ \BBA {} Simoes, P\BPBI J\BPBI A.%
\end{APACrefauthors}%
\unskip\
\newblock
\APACrefYearMonthDay{2018}{{\APACmonth{12}}}{}.
\newblock
{\BBOQ}\APACrefatitle {{Lyman Continuum Observations of Solar Flares Using
  SDO/EVE}} {{Lyman Continuum Observations of Solar Flares Using
  SDO/EVE}}.{\BBCQ}
\newblock
\APACjournalVolNumPages{Astrophysical Journal}{869}{1}{63}.
\PrintBackRefs{\CurrentBib}

\bibitem [\protect \citeauthoryear {%
{McClintock}%
, {Rottman}%
\BCBL {}\ \BBA {} {Woods}%
}{%
{McClintock}%
\ \protect \BOthers {.}}{%
{\protect \APACyear {2005}}%
}]{%
mccl05}
\APACinsertmetastar {%
mccl05}%
\begin{APACrefauthors}%
{McClintock}, W\BPBI E.%
, {Rottman}, G\BPBI J.%
\BCBL {}\ \BBA {} {Woods}, T\BPBI N.%
\end{APACrefauthors}%
\unskip\
\newblock
\APACrefYearMonthDay{2005}{{\APACmonth{08}}}{}.
\newblock
{\BBOQ}\APACrefatitle {{Solar-Stellar Irradiance Comparison Experiment II
  (Solstice II): Instrument Concept and Design}} {{Solar-Stellar Irradiance
  Comparison Experiment II (Solstice II): Instrument Concept and
  Design}}.{\BBCQ}
\newblock
\APACjournalVolNumPages{\solphys}{230}{}{225-258}.
\newblock
\begin{APACrefDOI} \doi{10.1007/s11207-005-7432-x} \end{APACrefDOI}
\PrintBackRefs{\CurrentBib}

\bibitem [\protect \citeauthoryear {%
Meier%
\ \BBA {} Prinz%
}{%
Meier%
\ \BBA {} Prinz%
}{%
{\protect \APACyear {1970}}%
}]{%
meie70}
\APACinsertmetastar {%
meie70}%
\begin{APACrefauthors}%
Meier, R\BPBI R.%
\BCBT {}\ \BBA {} Prinz, D\BPBI K.%
\end{APACrefauthors}%
\unskip\
\newblock
\APACrefYearMonthDay{1970}{{\APACmonth{12}}}{}.
\newblock
{\BBOQ}\APACrefatitle {{Absorption of the solar Lyman alpha line by geocoronal
  atomic hydrogen}} {{Absorption of the solar Lyman alpha line by geocoronal
  atomic hydrogen}}.{\BBCQ}
\newblock
\APACjournalVolNumPages{Journal of Geophysical Research}{75}{34}{6969--6979}.
\PrintBackRefs{\CurrentBib}

\bibitem [\protect \citeauthoryear {%
{Milligan}%
\ \BBA {} {Chamberlin}%
}{%
{Milligan}%
\ \BBA {} {Chamberlin}%
}{%
{\protect \APACyear {2016}}%
}]{%
mill16}
\APACinsertmetastar {%
mill16}%
\begin{APACrefauthors}%
{Milligan}, R\BPBI O.%
\BCBT {}\ \BBA {} {Chamberlin}, P\BPBI C.%
\end{APACrefauthors}%
\unskip\
\newblock
\APACrefYearMonthDay{2016}{{\APACmonth{03}}}{}.
\newblock
{\BBOQ}\APACrefatitle {{Anomalous temporal behaviour of broadband Ly{$\alpha$}
  observations during solar flares from SDO/EVE}} {{Anomalous temporal
  behaviour of broadband Ly{$\alpha$} observations during solar flares from
  SDO/EVE}}.{\BBCQ}
\newblock
\APACjournalVolNumPages{\aap}{587}{}{A123}.
\newblock
\begin{APACrefDOI} \doi{10.1051/0004-6361/201526682} \end{APACrefDOI}
\PrintBackRefs{\CurrentBib}

\bibitem [\protect \citeauthoryear {%
{Milligan}%
\ \protect \BOthers {.}}{%
{Milligan}%
\ \protect \BOthers {.}}{%
{\protect \APACyear {2012}}%
}]{%
mill12}
\APACinsertmetastar {%
mill12}%
\begin{APACrefauthors}%
{Milligan}, R\BPBI O.%
, {Chamberlin}, P\BPBI C.%
, {Hudson}, H\BPBI S.%
, {Woods}, T\BPBI N.%
, {Mathioudakis}, M.%
, {Fletcher}, L.%
\BDBL {}{Keenan}, F\BPBI P.%
\end{APACrefauthors}%
\unskip\
\newblock
\APACrefYearMonthDay{2012}{{\APACmonth{03}}}{}.
\newblock
{\BBOQ}\APACrefatitle {{Observations of Enhanced Extreme Ultraviolet Continua
  during an X-Class Solar Flare Using SDO/EVE}} {{Observations of Enhanced
  Extreme Ultraviolet Continua during an X-Class Solar Flare Using
  SDO/EVE}}.{\BBCQ}
\newblock
\APACjournalVolNumPages{\apjl}{748}{}{L14}.
\newblock
\begin{APACrefDOI} \doi{10.1088/2041-8205/748/1/L14} \end{APACrefDOI}
\PrintBackRefs{\CurrentBib}

\bibitem [\protect \citeauthoryear {%
Milligan%
, Fleck%
, Ireland%
, Fletcher%
\BCBL {}\ \BBA {} Dennis%
}{%
Milligan%
\ \protect \BOthers {.}}{%
{\protect \APACyear {2017}}%
}]{%
mill17}
\APACinsertmetastar {%
mill17}%
\begin{APACrefauthors}%
Milligan, R\BPBI O.%
, Fleck, B.%
, Ireland, J.%
, Fletcher, L.%
\BCBL {}\ \BBA {} Dennis, B\BPBI R.%
\end{APACrefauthors}%
\unskip\
\newblock
\APACrefYearMonthDay{2017}{{\APACmonth{10}}}{}.
\newblock
{\BBOQ}\APACrefatitle {{Detection of Three-minute Oscillations in Full-disk
  Lyalpha Emission during a Solar Flare}} {{Detection of Three-minute
  Oscillations in Full-disk Lyalpha Emission during a Solar Flare}}.{\BBCQ}
\newblock
\APACjournalVolNumPages{Astrophysical Journal}{848}{1}{L8}.
\PrintBackRefs{\CurrentBib}

\bibitem [\protect \citeauthoryear {%
Milligan%
\ \BBA {} Ireland%
}{%
Milligan%
\ \BBA {} Ireland%
}{%
{\protect \APACyear {2018}}%
}]{%
mill18}
\APACinsertmetastar {%
mill18}%
\begin{APACrefauthors}%
Milligan, R\BPBI O.%
\BCBT {}\ \BBA {} Ireland, J.%
\end{APACrefauthors}%
\unskip\
\newblock
\APACrefYearMonthDay{2018}{{\APACmonth{02}}}{}.
\newblock
{\BBOQ}\APACrefatitle {{On the Performance of Multi-Instrument Solar Flare
  Observations During Solar Cycle 24}} {{On the Performance of Multi-Instrument
  Solar Flare Observations During Solar Cycle 24}}.{\BBCQ}
\newblock
\APACjournalVolNumPages{Solar Physics}{293}{2}{411}.
\PrintBackRefs{\CurrentBib}

\bibitem [\protect \citeauthoryear {%
{Milligan}%
\ \protect \BOthers {.}}{%
{Milligan}%
\ \protect \BOthers {.}}{%
{\protect \APACyear {2014}}%
}]{%
mill14}
\APACinsertmetastar {%
mill14}%
\begin{APACrefauthors}%
{Milligan}, R\BPBI O.%
, {Kerr}, G\BPBI S.%
, {Dennis}, B\BPBI R.%
, {Hudson}, H\BPBI S.%
, {Fletcher}, L.%
, {Allred}, J\BPBI C.%
\BDBL {}{Keenan}, F\BPBI P.%
\end{APACrefauthors}%
\unskip\
\newblock
\APACrefYearMonthDay{2014}{{\APACmonth{10}}}{}.
\newblock
{\BBOQ}\APACrefatitle {{The Radiated Energy Budget of Chromospheric Plasma in a
  Major Solar Flare Deduced from Multi-wavelength Observations}} {{The Radiated
  Energy Budget of Chromospheric Plasma in a Major Solar Flare Deduced from
  Multi-wavelength Observations}}.{\BBCQ}
\newblock
\APACjournalVolNumPages{\apj}{793}{}{70}.
\newblock
\begin{APACrefDOI} \doi{10.1088/0004-637X/793/2/70} \end{APACrefDOI}
\PrintBackRefs{\CurrentBib}

\bibitem [\protect \citeauthoryear {%
Nusinov%
\ \BBA {} Kazachevskaya%
}{%
Nusinov%
\ \BBA {} Kazachevskaya%
}{%
{\protect \APACyear {2006}}%
}]{%
nusi06}
\APACinsertmetastar {%
nusi06}%
\begin{APACrefauthors}%
Nusinov, A\BPBI A.%
\BCBT {}\ \BBA {} Kazachevskaya, T\BPBI V.%
\end{APACrefauthors}%
\unskip\
\newblock
\APACrefYearMonthDay{2006}{{\APACmonth{03}}}{}.
\newblock
{\BBOQ}\APACrefatitle {{Extreme ultraviolet and X-ray emission of solar flares
  as observed from the CORONAS-F spacecraft in 2001{\textendash}2003}}
  {{Extreme ultraviolet and X-ray emission of solar flares as observed from the
  CORONAS-F spacecraft in 2001{\textendash}2003}}.{\BBCQ}
\newblock
\APACjournalVolNumPages{Solar System Research}{40}{2}{111--116}.
\PrintBackRefs{\CurrentBib}

\bibitem [\protect \citeauthoryear {%
{Parnell}%
\ \BBA {} {Jupp}%
}{%
{Parnell}%
\ \BBA {} {Jupp}%
}{%
{\protect \APACyear {2000}}%
}]{%
parn00}
\APACinsertmetastar {%
parn00}%
\begin{APACrefauthors}%
{Parnell}, C\BPBI E.%
\BCBT {}\ \BBA {} {Jupp}, P\BPBI E.%
\end{APACrefauthors}%
\unskip\
\newblock
\APACrefYearMonthDay{2000}{{\APACmonth{01}}}{}.
\newblock
{\BBOQ}\APACrefatitle {{Statistical Analysis of the Energy Distribution of
  Nanoflares in the Quiet Sun}} {{Statistical Analysis of the Energy
  Distribution of Nanoflares in the Quiet Sun}}.{\BBCQ}
\newblock
\APACjournalVolNumPages{\apj}{529}{}{554-569}.
\newblock
\begin{APACrefDOI} \doi{10.1086/308271} \end{APACrefDOI}
\PrintBackRefs{\CurrentBib}

\bibitem [\protect \citeauthoryear {%
{Pesnell}%
, {Thompson}%
\BCBL {}\ \BBA {} {Chamberlin}%
}{%
{Pesnell}%
\ \protect \BOthers {.}}{%
{\protect \APACyear {2012}}%
}]{%
pesn12}
\APACinsertmetastar {%
pesn12}%
\begin{APACrefauthors}%
{Pesnell}, W\BPBI D.%
, {Thompson}, B\BPBI J.%
\BCBL {}\ \BBA {} {Chamberlin}, P\BPBI C.%
\end{APACrefauthors}%
\unskip\
\newblock
\APACrefYearMonthDay{2012}{{\APACmonth{01}}}{}.
\newblock
{\BBOQ}\APACrefatitle {{The Solar Dynamics Observatory (SDO)}} {{The Solar
  Dynamics Observatory (SDO)}}.{\BBCQ}
\newblock
\APACjournalVolNumPages{\solphys}{275}{}{3-15}.
\newblock
\begin{APACrefDOI} \doi{10.1007/s11207-011-9841-3} \end{APACrefDOI}
\PrintBackRefs{\CurrentBib}

\bibitem [\protect \citeauthoryear {%
Raulin%
\ \protect \BOthers {.}}{%
Raulin%
\ \protect \BOthers {.}}{%
{\protect \APACyear {2013}}%
}]{%
raul13}
\APACinsertmetastar {%
raul13}%
\begin{APACrefauthors}%
Raulin, J\BHBI P.%
, Trottet, G.%
, Kretzschmar, M.%
, Macotela, E\BPBI L.%
, Pacini, A.%
, Bertoni, F\BPBI C\BPBI P.%
\BCBL {}\ \BBA {} Dammasch, I\BPBI E.%
\end{APACrefauthors}%
\unskip\
\newblock
\APACrefYearMonthDay{2013}{{\APACmonth{01}}}{}.
\newblock
{\BBOQ}\APACrefatitle {{Response of the low ionosphere to X-ray and
  Lyman-$\alpha$ solar flare emissions}} {{Response of the low ionosphere to
  X-ray and Lyman-$\alpha$ solar flare emissions}}.{\BBCQ}
\newblock
\APACjournalVolNumPages{Journal of Geophysical Research: Space
  Physics}{118}{1}{570--575}.
\PrintBackRefs{\CurrentBib}

\bibitem [\protect \citeauthoryear {%
Rochus%
, Auch{\`e}re%
, Berghmans%
\BCBL {}\ \BBA {} al%
}{%
Rochus%
\ \protect \BOthers {.}}{%
{\protect \APACyear {2020}}%
}]{%
roch20}
\APACinsertmetastar {%
roch20}%
\begin{APACrefauthors}%
Rochus, P.%
, Auch{\`e}re, F.%
, Berghmans, D.%
\BCBL {}\ \BBA {} al, e.%
\end{APACrefauthors}%
\unskip\
\newblock
\APACrefYearMonthDay{2020}{{\APACmonth{01}}}{}.
\newblock
{\BBOQ}\APACrefatitle {{The Solar Orbiter EUI instrument: The Extreme
  Ultraviolet Imager}} {{The Solar Orbiter EUI instrument: The Extreme
  Ultraviolet Imager}}.{\BBCQ}
\newblock
\APACjournalVolNumPages{Astronomy and Astrophysics}{}{}{}.
\PrintBackRefs{\CurrentBib}

\bibitem [\protect \citeauthoryear {%
Ryan%
, Dominique%
, Seaton%
, Stegen%
\BCBL {}\ \BBA {} White%
}{%
Ryan%
\ \protect \BOthers {.}}{%
{\protect \APACyear {2016}}%
}]{%
ryan16}
\APACinsertmetastar {%
ryan16}%
\begin{APACrefauthors}%
Ryan, D\BPBI F.%
, Dominique, M.%
, Seaton, D.%
, Stegen, K.%
\BCBL {}\ \BBA {} White, A.%
\end{APACrefauthors}%
\unskip\
\newblock
\APACrefYearMonthDay{2016}{{\APACmonth{08}}}{}.
\newblock
{\BBOQ}\APACrefatitle {{Effects of flare definitions on the statistics of
  derived flare distributions}} {{Effects of flare definitions on the
  statistics of derived flare distributions}}.{\BBCQ}
\newblock
\APACjournalVolNumPages{Astronomy and Astrophysics}{592}{}{A133}.
\PrintBackRefs{\CurrentBib}

\bibitem [\protect \citeauthoryear {%
{Ryan}%
\ \protect \BOthers {.}}{%
{Ryan}%
\ \protect \BOthers {.}}{%
{\protect \APACyear {2012}}%
}]{%
ryan12}
\APACinsertmetastar {%
ryan12}%
\begin{APACrefauthors}%
{Ryan}, D\BPBI F.%
, {Milligan}, R\BPBI O.%
, {Gallagher}, P\BPBI T.%
, {Dennis}, B\BPBI R.%
, {Tolbert}, A\BPBI K.%
, {Schwartz}, R\BPBI A.%
\BCBL {}\ \BBA {} {Young}, C\BPBI A.%
\end{APACrefauthors}%
\unskip\
\newblock
\APACrefYearMonthDay{2012}{{\APACmonth{10}}}{}.
\newblock
{\BBOQ}\APACrefatitle {{The Thermal Properties of Solar Flares over Three Solar
  Cycles Using GOES X-Ray Observations}} {{The Thermal Properties of Solar
  Flares over Three Solar Cycles Using GOES X-Ray Observations}}.{\BBCQ}
\newblock
\APACjournalVolNumPages{\apjs}{202}{}{11}.
\newblock
\begin{APACrefDOI} \doi{10.1088/0067-0049/202/2/11} \end{APACrefDOI}
\PrintBackRefs{\CurrentBib}

\bibitem [\protect \citeauthoryear {%
{Schrijver}%
\ \protect \BOthers {.}}{%
{Schrijver}%
\ \protect \BOthers {.}}{%
{\protect \APACyear {2012}}%
}]{%
schr12}
\APACinsertmetastar {%
schr12}%
\begin{APACrefauthors}%
{Schrijver}, C\BPBI J.%
, {Beer}, J.%
, {Baltensperger}, U.%
, {Cliver}, E\BPBI W.%
, {G{\"u}del}, M.%
, {Hudson}, H\BPBI S.%
\BDBL {}{Wolff}, E\BPBI W.%
\end{APACrefauthors}%
\unskip\
\newblock
\APACrefYearMonthDay{2012}{{\APACmonth{08}}}{}.
\newblock
{\BBOQ}\APACrefatitle {{Estimating the frequency of extremely energetic solar
  events, based on solar, stellar, lunar, and terrestrial records}}
  {{Estimating the frequency of extremely energetic solar events, based on
  solar, stellar, lunar, and terrestrial records}}.{\BBCQ}
\newblock
\APACjournalVolNumPages{Journal of Geophysical Research (Space
  Physics)}{117}{A8}{A08103}.
\newblock
\begin{APACrefDOI} \doi{10.1029/2012JA017706} \end{APACrefDOI}
\PrintBackRefs{\CurrentBib}

\bibitem [\protect \citeauthoryear {%
Sch{\"u}hle%
, Halain%
, Meining%
\BCBL {}\ \BBA {} Teriaca%
}{%
Sch{\"u}hle%
\ \protect \BOthers {.}}{%
{\protect \APACyear {2011}}%
}]{%
schu11}
\APACinsertmetastar {%
schu11}%
\begin{APACrefauthors}%
Sch{\"u}hle, U.%
, Halain, J\BHBI P.%
, Meining, S.%
\BCBL {}\ \BBA {} Teriaca, L.%
\end{APACrefauthors}%
\unskip\
\newblock
\APACrefYearMonthDay{2011}{{\APACmonth{09}}}{}.
\newblock
{\BBOQ}\APACrefatitle {{The Lyman-alpha telescope of the extreme ultraviolet
  imager on Solar Orbiter}} {{The Lyman-alpha telescope of the extreme
  ultraviolet imager on Solar Orbiter}}.{\BBCQ}
\newblock
\BIn{} S.~Fineschi\ \BBA {} J.~Fennelly\ (\BEDS), \APACrefbtitle {SPIE Optical
  Engineering + Applications} {Spie optical engineering + applications}\
  (\BPGS\ 81480K--81480K--11).
\newblock
\APACaddressPublisher{}{SPIE}.
\PrintBackRefs{\CurrentBib}

\bibitem [\protect \citeauthoryear {%
{Siskind}%
, {Barth}%
\BCBL {}\ \BBA {} {Cleary}%
}{%
{Siskind}%
\ \protect \BOthers {.}}{%
{\protect \APACyear {1990}}%
}]{%
sisk90}
\APACinsertmetastar {%
sisk90}%
\begin{APACrefauthors}%
{Siskind}, D\BPBI E.%
, {Barth}, C\BPBI A.%
\BCBL {}\ \BBA {} {Cleary}, D\BPBI D.%
\end{APACrefauthors}%
\unskip\
\newblock
\APACrefYearMonthDay{1990}{Apr}{}.
\newblock
{\BBOQ}\APACrefatitle {{The possible effect of solar soft X rays on
  thermospheric nitric oxide}} {{The possible effect of solar soft X rays on
  thermospheric nitric oxide}}.{\BBCQ}
\newblock
\APACjournalVolNumPages{\jgr}{95}{A4}{4311-4317}.
\newblock
\begin{APACrefDOI} \doi{10.1029/JA095iA04p04311} \end{APACrefDOI}
\PrintBackRefs{\CurrentBib}

\bibitem [\protect \citeauthoryear {%
{Siskind}%
, {Strickland}%
, {Meier}%
, {Majeed}%
\BCBL {}\ \BBA {} {Eparvier}%
}{%
{Siskind}%
\ \protect \BOthers {.}}{%
{\protect \APACyear {1995}}%
}]{%
sisk95}
\APACinsertmetastar {%
sisk95}%
\begin{APACrefauthors}%
{Siskind}, D\BPBI E.%
, {Strickland}, D\BPBI J.%
, {Meier}, R\BPBI R.%
, {Majeed}, T.%
\BCBL {}\ \BBA {} {Eparvier}, F\BPBI G.%
\end{APACrefauthors}%
\unskip\
\newblock
\APACrefYearMonthDay{1995}{Oct}{}.
\newblock
{\BBOQ}\APACrefatitle {{On the relationship between the solar soft X ray flux
  and thermospheric nitric oxide: An update with an improved photoelectron
  model}} {{On the relationship between the solar soft X ray flux and
  thermospheric nitric oxide: An update with an improved photoelectron
  model}}.{\BBCQ}
\newblock
\APACjournalVolNumPages{\jgr}{100}{A10}{19687-19694}.
\newblock
\begin{APACrefDOI} \doi{10.1029/95JA01609} \end{APACrefDOI}
\PrintBackRefs{\CurrentBib}

\bibitem [\protect \citeauthoryear {%
{Stewart}%
}{%
{Stewart}%
}{%
{\protect \APACyear {1861}}%
}]{%
stew61}
\APACinsertmetastar {%
stew61}%
\begin{APACrefauthors}%
{Stewart}, B.%
\end{APACrefauthors}%
\unskip\
\newblock
\APACrefYearMonthDay{1861}{Jan}{}.
\newblock
{\BBOQ}\APACrefatitle {{On the Great Magnetic Disturbance Which Extended from
  August 28 to September 7, 1859, as Recorded by Photography at the Kew
  Observatory}} {{On the Great Magnetic Disturbance Which Extended from August
  28 to September 7, 1859, as Recorded by Photography at the Kew
  Observatory}}.{\BBCQ}
\newblock
\APACjournalVolNumPages{Philosophical Transactions of the Royal Society of
  London Series I}{151}{}{423-430}.
\PrintBackRefs{\CurrentBib}

\bibitem [\protect \citeauthoryear {%
Teriaca%
\ \protect \BOthers {.}}{%
Teriaca%
\ \protect \BOthers {.}}{%
{\protect \APACyear {2011}}%
}]{%
teri11}
\APACinsertmetastar {%
teri11}%
\begin{APACrefauthors}%
Teriaca, L.%
, Andretta, V.%
, Auch{\`e}re, F.%
, Brown, C\BPBI M.%
, Buchlin, E.%
, Cauzzi, G.%
\BDBL {}Young, P.%
\end{APACrefauthors}%
\unskip\
\newblock
\APACrefYearMonthDay{2011}{{\APACmonth{09}}}{}.
\newblock
{\BBOQ}\APACrefatitle {{LEMUR: Large European Module for solar Ultraviolet
  Research. European contribution to JAXA's Solar-C mission}} {{LEMUR: Large
  European Module for solar Ultraviolet Research. European contribution to
  JAXA's Solar-C mission}}.{\BBCQ}
\newblock
\APACjournalVolNumPages{Experimental Astronomy}{}{2}{273--309}.
\PrintBackRefs{\CurrentBib}

\bibitem [\protect \citeauthoryear {%
{Thomson}%
\ \BBA {} {Clilverd}%
}{%
{Thomson}%
\ \BBA {} {Clilverd}%
}{%
{\protect \APACyear {2001}}%
}]{%
thomson_clivard}
\APACinsertmetastar {%
thomson_clivard}%
\begin{APACrefauthors}%
{Thomson}, N\BPBI R.%
\BCBT {}\ \BBA {} {Clilverd}, M\BPBI A.%
\end{APACrefauthors}%
\unskip\
\newblock
\APACrefYearMonthDay{2001}{{\APACmonth{11}}}{}.
\newblock
{\BBOQ}\APACrefatitle {{Solar flare induced ionospheric D-region enhancements
  from VLF amplitude observations}} {{Solar flare induced ionospheric D-region
  enhancements from VLF amplitude observations}}.{\BBCQ}
\newblock
\APACjournalVolNumPages{Journal of Atmospheric and Solar-Terrestrial
  Physics}{63}{16}{1729-1737}.
\newblock
\begin{APACrefDOI} \doi{10.1016/S1364-6826(01)00048-7} \end{APACrefDOI}
\PrintBackRefs{\CurrentBib}

\bibitem [\protect \citeauthoryear {%
{Thomson}%
, {Rodger}%
\BCBL {}\ \BBA {} {Clilverd}%
}{%
{Thomson}%
\ \protect \BOthers {.}}{%
{\protect \APACyear {2005}}%
}]{%
thomson2011}
\APACinsertmetastar {%
thomson2011}%
\begin{APACrefauthors}%
{Thomson}, N\BPBI R.%
, {Rodger}, C\BPBI J.%
\BCBL {}\ \BBA {} {Clilverd}, M\BPBI A.%
\end{APACrefauthors}%
\unskip\
\newblock
\APACrefYearMonthDay{2005}{{\APACmonth{06}}}{}.
\newblock
{\BBOQ}\APACrefatitle {{Large solar flares and their ionospheric D region
  enhancements}} {{Large solar flares and their ionospheric D region
  enhancements}}.{\BBCQ}
\newblock
\APACjournalVolNumPages{Journal of Geophysical Research (Space
  Physics)}{110}{A6}{A06306}.
\newblock
\begin{APACrefDOI} \doi{10.1029/2005JA011008} \end{APACrefDOI}
\PrintBackRefs{\CurrentBib}

\bibitem [\protect \citeauthoryear {%
{Torrence}%
\ \BBA {} {Compo}%
}{%
{Torrence}%
\ \BBA {} {Compo}%
}{%
{\protect \APACyear {1998}}%
}]{%
torr98}
\APACinsertmetastar {%
torr98}%
\begin{APACrefauthors}%
{Torrence}, C.%
\BCBT {}\ \BBA {} {Compo}, G\BPBI P.%
\end{APACrefauthors}%
\unskip\
\newblock
\APACrefYearMonthDay{1998}{{\APACmonth{01}}}{}.
\newblock
{\BBOQ}\APACrefatitle {{A Practical Guide to Wavelet Analysis.}} {{A Practical
  Guide to Wavelet Analysis.}}{\BBCQ}
\newblock
\APACjournalVolNumPages{Bulletin of the American Meteorological
  Society}{79}{}{61-78}.
\newblock
\begin{APACrefDOI} \doi{10.1175/1520-0477(1998)079<0061:APGTWA>2.0.CO;2}
  \end{APACrefDOI}
\PrintBackRefs{\CurrentBib}

\bibitem [\protect \citeauthoryear {%
{Tranquille}%
, {Hurley}%
\BCBL {}\ \BBA {} {Hudson}%
}{%
{Tranquille}%
\ \protect \BOthers {.}}{%
{\protect \APACyear {2009}}%
}]{%
tran09}
\APACinsertmetastar {%
tran09}%
\begin{APACrefauthors}%
{Tranquille}, C.%
, {Hurley}, K.%
\BCBL {}\ \BBA {} {Hudson}, H\BPBI S.%
\end{APACrefauthors}%
\unskip\
\newblock
\APACrefYearMonthDay{2009}{{\APACmonth{08}}}{}.
\newblock
{\BBOQ}\APACrefatitle {{The Ulysses Catalog of Solar Hard X-Ray Flares}} {{The
  Ulysses Catalog of Solar Hard X-Ray Flares}}.{\BBCQ}
\newblock
\APACjournalVolNumPages{\solphys}{258}{}{141-166}.
\newblock
\begin{APACrefDOI} \doi{10.1007/s11207-009-9387-9} \end{APACrefDOI}
\PrintBackRefs{\CurrentBib}

\bibitem [\protect \citeauthoryear {%
Veronig%
, Temmer%
, Hanslmeier%
, Otruba%
\BCBL {}\ \BBA {} Messerotti%
}{%
Veronig%
\ \protect \BOthers {.}}{%
{\protect \APACyear {2002}}%
}]{%
vero02}
\APACinsertmetastar {%
vero02}%
\begin{APACrefauthors}%
Veronig, A.%
, Temmer, M.%
, Hanslmeier, A.%
, Otruba, W.%
\BCBL {}\ \BBA {} Messerotti, M.%
\end{APACrefauthors}%
\unskip\
\newblock
\APACrefYearMonthDay{2002}{{\APACmonth{02}}}{}.
\newblock
{\BBOQ}\APACrefatitle {{Temporal aspects and frequency distributions of solar
  soft X-ray flares}} {{Temporal aspects and frequency distributions of solar
  soft X-ray flares}}.{\BBCQ}
\newblock
\APACjournalVolNumPages{Astronomy and Astrophysics}{382}{3}{1070--1080}.
\PrintBackRefs{\CurrentBib}

\bibitem [\protect \citeauthoryear {%
Viereck%
\ \protect \BOthers {.}}{%
Viereck%
\ \protect \BOthers {.}}{%
{\protect \APACyear {2007}}%
}]{%
vier07}
\APACinsertmetastar {%
vier07}%
\begin{APACrefauthors}%
Viereck, R.%
, Hanser, F.%
, Wise, J.%
, Guha, S.%
, Jones, A.%
, McMullin, D.%
\BDBL {}Evans, S.%
\end{APACrefauthors}%
\unskip\
\newblock
\APACrefYearMonthDay{2007}{{\APACmonth{09}}}{}.
\newblock
{\BBOQ}\APACrefatitle {{Solar extreme ultraviolet irradiance observations from
  GOES: design characteristics and initial performance}} {{Solar extreme
  ultraviolet irradiance observations from GOES: design characteristics and
  initial performance}}.{\BBCQ}
\newblock
\BIn{} S.~Fineschi\ \BBA {} R\BPBI A.~Viereck\ (\BEDS), \APACrefbtitle {Solar
  Physics and Space Weather Instrumentation II} {Solar physics and space
  weather instrumentation ii}\ (\BPG~66890K).
\newblock
\APACaddressPublisher{}{International Society for Optics and Photonics}.
\PrintBackRefs{\CurrentBib}

\bibitem [\protect \citeauthoryear {%
White%
, Thomas%
\BCBL {}\ \BBA {} Schwartz%
}{%
White%
\ \protect \BOthers {.}}{%
{\protect \APACyear {2005}}%
}]{%
whit05}
\APACinsertmetastar {%
whit05}%
\begin{APACrefauthors}%
White, S\BPBI M.%
, Thomas, R\BPBI J.%
\BCBL {}\ \BBA {} Schwartz, R\BPBI A.%
\end{APACrefauthors}%
\unskip\
\newblock
\APACrefYearMonthDay{2005}{{\APACmonth{04}}}{}.
\newblock
{\BBOQ}\APACrefatitle {{Updated Expressions for Determining Temperatures and
  Emission Measures from Goes Soft X-Ray Measurements}} {{Updated Expressions
  for Determining Temperatures and Emission Measures from Goes Soft X-Ray
  Measurements}}.{\BBCQ}
\newblock
\APACjournalVolNumPages{Solar Physics}{227}{2}{231--248}.
\PrintBackRefs{\CurrentBib}

\bibitem [\protect \citeauthoryear {%
Wilhelm%
\ \protect \BOthers {.}}{%
Wilhelm%
\ \protect \BOthers {.}}{%
{\protect \APACyear {1995}}%
}]{%
wilh95}
\APACinsertmetastar {%
wilh95}%
\begin{APACrefauthors}%
Wilhelm, K.%
, Curdt, W.%
, Marsch, E.%
, Sch{\"u}hle, U.%
, Lemaire, P.%
, Gabriel, A.%
\BDBL {}Siegmund, O\BPBI H\BPBI W.%
\end{APACrefauthors}%
\unskip\
\newblock
\APACrefYearMonthDay{1995}{{\APACmonth{12}}}{}.
\newblock
{\BBOQ}\APACrefatitle {{SUMER - Solar Ultraviolet Measurements of Emitted
  Radiation}} {{SUMER - Solar Ultraviolet Measurements of Emitted
  Radiation}}.{\BBCQ}
\newblock
\APACjournalVolNumPages{Solar Physics}{162}{1}{189--231}.
\PrintBackRefs{\CurrentBib}

\bibitem [\protect \citeauthoryear {%
Woods%
}{%
Woods%
}{%
{\protect \APACyear {2008}}%
}]{%
wood08}
\APACinsertmetastar {%
wood08}%
\begin{APACrefauthors}%
Woods, T\BPBI N.%
\end{APACrefauthors}%
\unskip\
\newblock
\APACrefYearMonthDay{2008}{{\APACmonth{09}}}{}.
\newblock
{\BBOQ}\APACrefatitle {{Recent advances in observations and modeling of the
  solar ultraviolet and X-ray spectral irradiance}} {{Recent advances in
  observations and modeling of the solar ultraviolet and X-ray spectral
  irradiance}}.{\BBCQ}
\newblock
\APACjournalVolNumPages{Advances in Space Research}{42}{}{895}.
\PrintBackRefs{\CurrentBib}

\bibitem [\protect \citeauthoryear {%
{Woods}%
\ \protect \BOthers {.}}{%
{Woods}%
\ \protect \BOthers {.}}{%
{\protect \APACyear {2009}}%
}]{%
wood09}
\APACinsertmetastar {%
wood09}%
\begin{APACrefauthors}%
{Woods}, T\BPBI N.%
, {Chamberlin}, P\BPBI C.%
, {Harder}, J\BPBI W.%
, {Hock}, R\BPBI A.%
, {Snow}, M.%
, {Eparvier}, F\BPBI G.%
\BDBL {}{Richard}, E\BPBI C.%
\end{APACrefauthors}%
\unskip\
\newblock
\APACrefYearMonthDay{2009}{Jan}{}.
\newblock
{\BBOQ}\APACrefatitle {{Solar Irradiance Reference Spectra (SIRS) for the 2008
  Whole Heliosphere Interval (WHI)}} {{Solar Irradiance Reference Spectra
  (SIRS) for the 2008 Whole Heliosphere Interval (WHI)}}.{\BBCQ}
\newblock
\APACjournalVolNumPages{\grl}{36}{1}{L01101}.
\newblock
\begin{APACrefDOI} \doi{10.1029/2008GL036373} \end{APACrefDOI}
\PrintBackRefs{\CurrentBib}

\bibitem [\protect \citeauthoryear {%
{Woods}%
\ \protect \BOthers {.}}{%
{Woods}%
\ \protect \BOthers {.}}{%
{\protect \APACyear {2004}}%
}]{%
wood04}
\APACinsertmetastar {%
wood04}%
\begin{APACrefauthors}%
{Woods}, T\BPBI N.%
, {Eparvier}, F\BPBI G.%
, {Fontenla}, J.%
, {Harder}, J.%
, {Kopp}, G.%
, {McClintock}, W\BPBI E.%
\BDBL {}{Snow}, M.%
\end{APACrefauthors}%
\unskip\
\newblock
\APACrefYearMonthDay{2004}{{\APACmonth{05}}}{}.
\newblock
{\BBOQ}\APACrefatitle {{Solar irradiance variability during the October 2003
  solar storm period}} {{Solar irradiance variability during the October 2003
  solar storm period}}.{\BBCQ}
\newblock
\APACjournalVolNumPages{\grl}{31}{}{L10802}.
\newblock
\begin{APACrefDOI} \doi{10.1029/2004GL019571} \end{APACrefDOI}
\PrintBackRefs{\CurrentBib}

\bibitem [\protect \citeauthoryear {%
{Woods}%
\ \protect \BOthers {.}}{%
{Woods}%
\ \protect \BOthers {.}}{%
{\protect \APACyear {2012}}%
}]{%
wood12}
\APACinsertmetastar {%
wood12}%
\begin{APACrefauthors}%
{Woods}, T\BPBI N.%
, {Eparvier}, F\BPBI G.%
, {Hock}, R.%
, {Jones}, A\BPBI R.%
, {Woodraska}, D.%
, {Judge}, D.%
\BDBL {}{Viereck}, R.%
\end{APACrefauthors}%
\unskip\
\newblock
\APACrefYearMonthDay{2012}{{\APACmonth{01}}}{}.
\newblock
{\BBOQ}\APACrefatitle {{Extreme Ultraviolet Variability Experiment (EVE) on the
  Solar Dynamics Observatory (SDO): Overview of Science Objectives, Instrument
  Design, Data Products, and Model Developments}} {{Extreme Ultraviolet
  Variability Experiment (EVE) on the Solar Dynamics Observatory (SDO):
  Overview of Science Objectives, Instrument Design, Data Products, and Model
  Developments}}.{\BBCQ}
\newblock
\APACjournalVolNumPages{\solphys}{275}{}{115-143}.
\newblock
\begin{APACrefDOI} \doi{10.1007/s11207-009-9487-6} \end{APACrefDOI}
\PrintBackRefs{\CurrentBib}

\bibitem [\protect \citeauthoryear {%
{Woods}%
, {Kopp}%
\BCBL {}\ \BBA {} {Chamberlin}%
}{%
{Woods}%
\ \protect \BOthers {.}}{%
{\protect \APACyear {2006}}%
}]{%
wood06}
\APACinsertmetastar {%
wood06}%
\begin{APACrefauthors}%
{Woods}, T\BPBI N.%
, {Kopp}, G.%
\BCBL {}\ \BBA {} {Chamberlin}, P\BPBI C.%
\end{APACrefauthors}%
\unskip\
\newblock
\APACrefYearMonthDay{2006}{{\APACmonth{10}}}{}.
\newblock
{\BBOQ}\APACrefatitle {{Contributions of the solar ultraviolet irradiance to
  the total solar irradiance during large flares}} {{Contributions of the solar
  ultraviolet irradiance to the total solar irradiance during large
  flares}}.{\BBCQ}
\newblock
\APACjournalVolNumPages{Journal of Geophysical Research (Space
  Physics)}{111}{}{A10S14}.
\newblock
\begin{APACrefDOI} \doi{10.1029/2005JA011507} \end{APACrefDOI}
\PrintBackRefs{\CurrentBib}

\bibitem [\protect \citeauthoryear {%
Woods%
, Rottman%
, White%
, Fontenla%
\BCBL {}\ \BBA {} Avrett%
}{%
Woods%
\ \protect \BOthers {.}}{%
{\protect \APACyear {1995}}%
}]{%
wood95}
\APACinsertmetastar {%
wood95}%
\begin{APACrefauthors}%
Woods, T\BPBI N.%
, Rottman, G\BPBI J.%
, White, O\BPBI R.%
, Fontenla, J.%
\BCBL {}\ \BBA {} Avrett, E\BPBI H.%
\end{APACrefauthors}%
\unskip\
\newblock
\APACrefYearMonthDay{1995}{{\APACmonth{03}}}{}.
\newblock
{\BBOQ}\APACrefatitle {{The solar Ly-alpha line profile}} {{The solar Ly-alpha
  line profile}}.{\BBCQ}
\newblock
\APACjournalVolNumPages{The Astrophysical Journal}{442}{}{898--906}.
\PrintBackRefs{\CurrentBib}

\bibitem [\protect \citeauthoryear {%
{Woods}%
, {Tobiska}%
, {Rottman}%
\BCBL {}\ \BBA {} {Worden}%
}{%
{Woods}%
\ \protect \BOthers {.}}{%
{\protect \APACyear {2000}}%
}]{%
wood00}
\APACinsertmetastar {%
wood00}%
\begin{APACrefauthors}%
{Woods}, T\BPBI N.%
, {Tobiska}, W\BPBI K.%
, {Rottman}, G\BPBI J.%
\BCBL {}\ \BBA {} {Worden}, J\BPBI R.%
\end{APACrefauthors}%
\unskip\
\newblock
\APACrefYearMonthDay{2000}{{\APACmonth{12}}}{}.
\newblock
{\BBOQ}\APACrefatitle {{Improved solar Lyman {$\alpha$} irradiance modeling
  from 1947 through 1999 based on UARS observations}} {{Improved solar Lyman
  {$\alpha$} irradiance modeling from 1947 through 1999 based on UARS
  observations}}.{\BBCQ}
\newblock
\APACjournalVolNumPages{\jgr}{105}{}{27195-27216}.
\newblock
\begin{APACrefDOI} \doi{10.1029/2000JA000051} \end{APACrefDOI}
\PrintBackRefs{\CurrentBib}

\bibitem [\protect \citeauthoryear {%
{\v{Z}}igman%
, Grubor%
\BCBL {}\ \BBA {} {\v{S}}uli{\'c}%
}{%
{\v{Z}}igman%
\ \protect \BOthers {.}}{%
{\protect \APACyear {2007}}%
}]{%
zigm07}
\APACinsertmetastar {%
zigm07}%
\begin{APACrefauthors}%
{\v{Z}}igman, V.%
, Grubor, D.%
\BCBL {}\ \BBA {} {\v{S}}uli{\'c}, D.%
\end{APACrefauthors}%
\unskip\
\newblock
\APACrefYearMonthDay{2007}{}{}.
\newblock
{\BBOQ}\APACrefatitle {D-region electron density evaluated from VLF amplitude
  time delay during X-ray solar flares} {D-region electron density evaluated
  from vlf amplitude time delay during x-ray solar flares}.{\BBCQ}
\newblock
\APACjournalVolNumPages{Journal of atmospheric and solar-terrestrial
  physics}{69}{7}{775--792}.
\PrintBackRefs{\CurrentBib}

\end{thebibliography}
%
% don't specify bibliographystyle
%%%%%%%%%%%%%%%%%%%%%%%%%%%%%%%%%%%%%%%%%%%%%%%

% Please use ONLY \citet and \citep for reference citations.
% DO NOT use other cite commands (e.g., \cite, \citeyear, \nocite, \citealp, etc.).
%% Example \citet and \citep:
%  ...as shown by \citet{Boug10}, \citet{Buiz07}, \citet{Fra10},
%  \citet{Ghel00}, and \citet{Leit74}.

%  ...as shown by \citep{Boug10}, \citep{Buiz07}, \citep{Fra10},
%  \citep{Ghel00, Leit74}.

%  ...has been shown \citep [e.g.,][]{Boug10,Buiz07,Fra10}.

\end{document}